\documentclass[prb,aps,amsmath,amssymb,twocolumn]{revtex4}
\usepackage[dvipdfm]{graphicx}
\usepackage{dcolumn}
\usepackage{bm}
\usepackage{color}
\usepackage{ulem}

\begin{document}
\newcommand{\avrg}[1]{\left\langle #1 \right\rangle}
\newcommand{\eqsa}[1]{\begin{eqnarray} #1 \end{eqnarray}}
\newcommand{\eqwd}[1]{\begin{widetext}\begin{eqnarray} #1 \end{eqnarray}\end{widetext}}
\newcommand{\hatd}[2]{\hat{ #1 }^{\dagger}_{ #2 }}
\newcommand{\hatn}[2]{\hat{ #1 }^{\ }_{ #2 }}
\newcommand{\wdtd}[2]{\widetilde{ #1 }^{\dagger}_{ #2 }}
\newcommand{\wdtn}[2]{\widetilde{ #1 }^{\ }_{ #2 }}
\newcommand{\cond}[1]{\overline{ #1 }_{0}}
\newcommand{\conp}[2]{\overline{ #1 }_{0#2}}
\newcommand{\nn}{\nonumber\\}
\newcommand{\cdt}{$\cdot$}
\newcommand{\bra}[1]{\langle#1|}
\newcommand{\ket}[1]{|#1\rangle}
\newcommand{\braket}[2]{\langle #1 | #2 \rangle}
\newcommand{\bvec}[1]{\mbox{\boldmath$#1$}}
\newcommand{\blue}[1]{{#1}}
\newcommand{\bl}[1]{{#1}}
\newcommand{\bn}[1]{\textcolor{black}{#1}}
\newcommand{\rr}[1]{{#1}}
\newcommand{\bu}[1]{\textcolor{black}{#1}}
\newcommand{\red}[1]{{#1}}
\newcommand{\fj}[1]{{#1}}
\newcommand{\green}[1]{{#1}}
\newcommand{\gr}[1]{{#1}}
\newcommand{\tmag}[1]{\textcolor{black}{#1}}
\definecolor{green}{rgb}{0,0.5,0.1}
\definecolor{blue}{rgb}{0,0,0.8}
\preprint{APS/123-QED}

\title{
Metallic Interface Emerging at Magnetic Domain Wall of Antiferromagnetic Insulator 
\\ --- Fate of Extinct Weyl Electrons
}
\author{Youhei Yamaji$^{\ast}$ and Masatoshi Imada}
\affiliation{{\rm Department of Applied Physics, University of Tokyo, Hongo, Bunkyo-ku, Tokyo, 113-8656, Japan.
}}%
\begin{abstract}
Topological insulators, in contrast to ordinary semiconductors, accompany protected metallic surfaces described by Dirac-type fermions.
Here, we theoretically show another emergent two-dimensional metal embedded in the bulk insulator is realized at a magnetic domain wall. The domain wall
has \red{long been studied as ingredients of} both old-fashioned and leading-edge spintronics. The domain wall \red{here}, as an interface of seemingly trivial antiferromagnetic insulators, emergently realizes a functional interface preserved by zero modes with robust two-dimensional Fermi surfaces, where pyrochlore iridium oxides proposed to host condensed-matter realization of Weyl fermions offer such examples at low temperatures. The existence of ingap states pinned at domain walls, theoretically resembling spin/charge solitons in polyacetylene,
\textcolor{black}{and protected as the edge of
{\it hidden} one-dimensional weak Chern insulators
characterized by
a zero-dimensional class A topological
invariant,}
solves experimental puzzles observed in \red{$R_2$Ir$_2$O$_7$ with rare earth elements $R$.} The domain wall realizes a novel quantum confinement of electrons and embosses a net uniform magnetization, which enables magnetic control of electronic interface transports beyond semiconductor paradigm.
\end{abstract}
\maketitle
\section{Introduction}
Interfaces in semiconductor heterojunctions, field effect transistors,
and between vacuum and newly characterized topologically non-trivial semiconductors
host various two-dimensional electron systems tightly confined around these
interfaces,
which offer major playgrounds of electronics and spintronics.
Especially, topologically non-trivial semiconductors
classified as 
topological insulators
or Chern insulators\cite{HasanKane,RyuSchnyder}, in contrast to
usual band insulators, accompany protected metallic 
surfaces described by Dirac-type fermions\cite{Volkov1985,Pankratov1986,Fradkin1986}.
Another peculiar metallic state with truncated Fermi surface called ``arc" is predicted\cite{Wan11}
on the interfaces between vacuum and a newly recognized class of zero-gap semiconductor\cite{CastroNeto,Sun09}:
It hosts 
condensed-matter realization of Weyl fermions, initially proposed in
iridium pyrochlore oxides $R_2$Ir$_2$O$_7$
under a magnetic order 
\cite{Yanagishima01,Matsuhira11,
Tomiyasu12,Ueda12,Ishikawa,Matsuhira13}.  

In this article, we unveil that
magnetic domain walls
offer qualitatively novel interfaces
in magnetically ordered zero-gap semiconductors \red{such as $R_2$Ir$_2$O$_7$},
which are expected to host Weyl fermions in the bulk.
Magnetic domain walls have historically been of interest in both fundamental physics\cite{Kittel}
and technology\cite{Parkin}
as an archetypical and fundamental model for inhomogeneity originating from spontaneous symmetry breaking,
and, for example, as an essential ingredient for antique magnetic-bubble memory.
Recently, applications of spintronics, such as magnetic random access memories,
\tmag{have received renewed interest} in electric controls of magnetic domain walls.
We theoretically \red{show}, differently from these
aspects and applications of magnetic domain walls,
that  \red{a class of} magnetic domain walls 
\red{induces} unexpected interface metals accompanied by 
a net uniform magnetization, brought about by the insertion of the domain wall,
in the background
of seemingly trivial bulk antiferromagnetic insulator, where \red{uniform magnetization
is} cancelled each other \red{in the bulk}.
\tmag{The metallicity of the domain wall is triggered by the formation of Fermi arcs
at the domain walls, which originate from
the condesed-matter Weyl fermions, or the Weyl electrons, while
the Fermi arc evolves into the Fermi surface when the Weyl fermions are eliminated 
as detailed in this article.}

Robustness against perturbations and \red{the} anomalous electromagnetic responses
of Weyl fermions arising from the chiral anomaly
are the reasons why \red{the condensed-matter realization of Weyl fermions have been interested experimentally and theoretically}\cite{WilliamReview}.
The Weyl electrons \red{are, however,} easily annihilated in pair with Weyl electrons of the opposite chirality.
Consequently, the Fermi arc on the surface of the pyrochlore zero-gap semiconductor
survives 
only near the all-in/all-out-type antiferromagnetic transition temperature\cite{Ueda12,WK12}.
\begin{figure*}[htb]
\centering
\includegraphics[width=18.0cm]{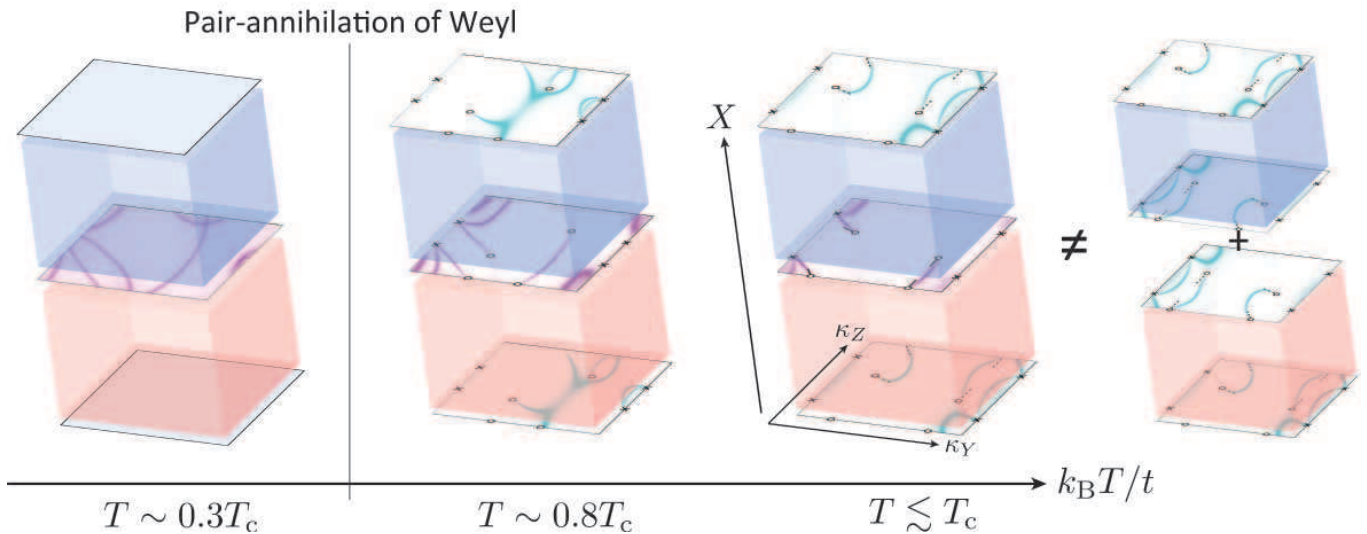}
\caption{
Two dimensional Brillouin zones illustrated for two different kinds of interfaces, namely,
surfaces of bulk crystals with magnetic orders (red or blue transparent cubes)
against vacua, and domain walls between two magnetic domains.
Evolution of Fermi arcs are shown for the surfaces between the bulk and vacua and for the magnetic domain walls between
two magnetic domains 
upon changes in temperatures $T$ (indicated by the horizontal axis).
\tmag{The evolution} occurs in a zero-gap semiconductors hosting Weyl electrons below its critical
temperatures for the magnetic order, $T_{\rm c}$.
The domains with two inequivalent magnetic orders that are mapped each other
through the time-reversal operation are illustrated as the red and blue transparent cubes
(called as ``all-out" and ``all-in" domains in the later discussion). 
These surfaces and domain walls are illustrated to be perpendicular to the $X$-axis.
The spectral functions at the Fermi level are shown in the momentum frame $(\kappa_Y, \kappa_Z)$ with finite broadening factor
for the sake of illustration. Here, stronger colors indicates higher spectral intensities.
In contrast to naive expectation that the magnetic domain walls simply consist of two surfaces between the bulk and vacua
(shown in the right end),
the magnetic domain wall offers a novel two-dimensional interface \tmag{distinct} from the surfaces:
Even after pair-annihilation of the Weyl electrons, the domain-wall metallic states remain and form a Fermi surface
(shown in the left end panel).
All of the Fermi arcs (and Fermi surface at lowest temperatures) are calculated by using the tight-binding hamiltonian Eq.(1)
with the $(01\overline{1})$-surfaces and domain walls introduced later, with the parameter set used
for Figs. \ref{fig5} and \ref{fig6}.
For detailed notations, see the captions of Figs.~\ref{fig5} and \ref{fig6}.
}
\label{fig1}
\end{figure*}

In contrast to
fragile Fermi arcs at surfaces,
here we show that
magnetic domain walls realize
metallic interfaces preserved by zero modes or ingap states with
robust
Fermi arc
or
Fermi surface even after pair-annihilation of Weyl electrons and even in
the seemingly trivial antiferromagnetic insulators, as summarized in Fig.~\ref{fig1}.
The zero modes follow a one-dimensional Dirac equation that protects ingap states.
\textcolor{black}{The existence of gapless mode at the domain wall is protected because
the bulk state is projected to a one-dimensional weak Chern insulator provided that
certain symmetries are satisfied. The domain wall can indeed satisfy these symmetries.}
\textcolor{black}{N}amely persisting metallicity pinned at domain walls
\textcolor{black}{is assumed.
The domain walls also}
maintain a ferromagnetic moment,
similarly to spin solitons in polyacetylene.
It may solve experimental puzzles of the
\textcolor{black}{iridium oxides}\textcolor{black}{, such as
bad insulating behaviors\cite{Yanagishima01,Matsuhira11}
with clear optical gaps\cite{Ueda12},
anomalous weak ferromagnetism\cite{Matsuhira11,Ishikawa},
and anomalous magneto-transports\cite{Matsuhira13,Disseler}
widely observed in the pyrochlore iridium oxides,
regardless of their detailed chemical compositions}.
Furthermore, it offers a novel quantum confinement of
electrons enabling magnetic control
of interface electronic transports.
For
quantum wells/crystal \red{grain boundaries},
location, orientation, and number of these interfaces can not
be controlled after their fabrications.
In contrast, our magnetic domain walls are tunable through
applied magnetic fields, and further host protected ingap
metallic domain-wall states.

\begin{figure}[htb]
\centering
\includegraphics[width=9.0cm]{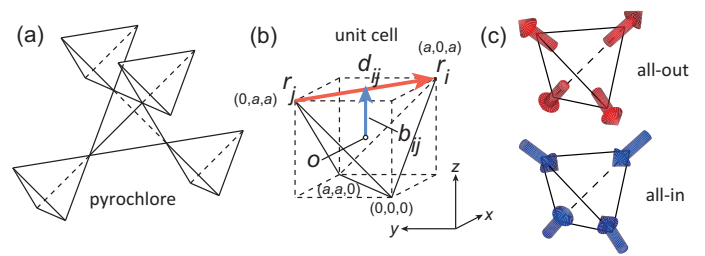}
\caption{
Magnetic structure and notations for spin-orbit interaction on 
pyrochlore lattice.
(a) Pyrochlore lattice structure,
(b) definition of effective spin-orbit couplings
on pyrochlore lattices,
and (c) all-in/all-out magnetic moment configuration. 
In (b), vectors $\vec{d}_{ij}$ and $\vec{b}_{ij}$
are illustrated for a specific bond.
The vector $\vec{d}_{ij}$ points from the $j$-th site $\vec{r}_{j}$ to the $i$-th site $\vec{r}_{i}$
and the vector $b_{ij}$ points from the center of the unit tetrahedron $O$
to the midpoint between $\vec{r}_{i}$ and $\vec{r}_{j}$.
\label{fig2}}
\end{figure}

Dirac/Weyl-type fermions\cite{Abrikosov71,LL9part2}
realized in 
crystalline solids with both strong spin-orbit couplings and Coulomb repulsion
are a subject of intensive studies\cite{CastroNeto,Pesin,Go,WK12,Moon12}.
We elucidate another prominent effects arising from combined interaction and topology  by
studying
single band Hubbard-type model on the pyrochlore
lattice (Fig.~\ref{fig2}(a)),
with the $J_{\rm eff}=1/2$-manifold
of the iridium pyrochlore oxides in mind\textcolor{black}{,
where $J_{\rm eff}$ is an effective total angular momentum
of 5$d$-orbitals of an iridium atom with five electrons:
We}
show that the Weyl electrons leave
behind their indelible trace with a Fermi surface at the magnetic domain
walls even after the pair-annihilations of them, namely
even when the Weyl electrons completely disappear and the bulk and surface
turn into an insulator.
This conclusion is supported by
fully unrestricted
Hartree-Fock calculations\textcolor{black}{, where
the self-consistent mean fields at every atoms
for charge density and three spin components
are fully relaxed,}
and Dirac equations for 
effective low-energy model.
Electronic states bound around the domain walls are formed,
whose origin is traced back to 
the bulk Weyl electrons \textcolor{black}{and their quantum chiral anomaly}.
The 
present domain-wall theory offers insights into a number of peculiar properties of $R_2$Ir$_2$O$_7$
including weak ferromagnetism with strong field dependence\cite{Ishikawa},
bad but stubborn electronic conduction\cite{Ueda12} and negative magnetoresistance\cite{Matsuhira13}. 

The outline of the present paper is as follows:
A theoretical model for pyrochlore iridium oxides is defined
in Sec. \ref{Section2}.
In Sec. \ref{Section3}, Weyl electrons and
their pair-annihilation in the present model
are described. We employ the standard $\vec{k}\cdot\vec{p}$
perturbation theory around the Fermi level, which generates 
the Luttinger hamiltonian as is derived in Appendix A and B. 
In Section \ref{Section4} together with Appendix D, 
the Luttinger hamiltonian is reduced to 
\textcolor{black}{one-dimensional} Dirac
\textcolor{black}{/three-dimensional Weyl} hamiltonians characterizing low-energy
electronic states around magnetic domain walls
(defined in Appendix \ref{ADWs})
that describe the bulk Weyl electrons, and Fermi arcs/surfaces
appearing at the domain walls.
Topological properties of the domain-wall states
are elucidated in Sec.\ref{newV} and Appendix E as well as in a simplified model 
in Sec. \ref{Section5}.
Similarity of the domain wall to the solitons in polyacetylene is discussed.
Unrestricted Hartree-Fock results on the model
are given in Sec. \ref{Section6}.
It is shown that the one-dimensional Dirac equation captures the essence of 
the full effective \textcolor{black}{tight-binding} model for the pyrochlore \textcolor{black}{iridium oxides}.
\textcolor{black}{Symmetric properties and symmetry-protected topological nature
of the domain-wall states are detailed in Sec. VIII.}
Relevance of our theoretical predictions to experimental
observation on pyrochlore iridium oxides is discussed
in Sec. \ref{Section7}. Bad insulating properties in the electronic conduction, 
and weak ferromagnetism, as well as large magnetoresistance in Nd and Gd compounds 
observed experimentally are naturally 
understood from the present theory. 

\section{Model of pyrochlore iridium oxides with spin-orbit interaction}
\label{Section2}
In this article, we employ \red{a simple} model describing essential physics
of iridium pyrochlore oxides, which is also \red{one of the} minimal models hosting bulk Weyl fermions:
The Hubbard hamiltonian with the onsite interaction $U$, transfer $t$ and spin-orbit coupling $\zeta$ decoded as
spin-dependent imaginary hopping
at the filling of one electron per site is introduced, 
\eqsa{
\hat{H} &=&-t\sum_{i,j}^{\rm n.n.}\sum_{\sigma}
        \left[
        \hatd{c}{i\sigma}\hatn{c}{j\sigma}
        +{\rm h.c.}
        \right]
 +U\sum_{i}\hat{n}_{i\uparrow}\hat{n}_{i\downarrow}
        \nn
        &&+i\zeta \sum_{i,j}^{\rm n.n.}
          \sum_{\alpha,\beta=\uparrow,\downarrow}
          \hatd{c}{i\alpha}
          \left(\vec{\hat{\sigma}}\cdot
          \frac{\vec{b}_{ij}\times\vec{d}_{ij}}{|\vec{b}_{ij}\times\vec{d}_{ij}|}\right)_{\alpha\beta}
          \hatn{c}{j\beta},
\label{TB}
}
where a fermionic operator $\hatd{c}{i\sigma}$ ($\hatn{c}{i\sigma}$) creates (annihilates) an electron
with $\sigma$-spin at $i$-th site.
\red{Here, the effective spin-orbit coupling described by the $\zeta$-term is given by pseudovectors $\vec{b}_{ij}\times\vec{d}_{ij}$ 
illustrated in Fig. \ref{fig2}(b).
This is the unique form of the spin-orbit interaction as the nearest-neighbor hopping matrices allowed by the time-reversal symmetry and the point-group symmetry
of the pyrochlore lattice,
except rotations of the global spin quantization axis.}

The sign of $\zeta$ determines 
the electronic structure of the model\cite{Guo09,Kurita11} and its ground state magnetism:
The zero-gap semiconductors are realized for $\zeta<0$
while, for $\zeta > 0$, the system becomes a topological insulator 
in the absence of interaction.
The magnetic ground state for $\zeta <0$ and $U > 0$ is
the all-in/all-out order (Fig.~\ref{fig2}(c)), where the
magnetic moment at each site points away from
or toward
the center of the tetrahedron and feels Ising-type anisotropy\cite{Elhajal}.
Thus, there remains
two-fold 
degeneracy of the order.
\begin{figure*}[htb]
\centering
\includegraphics[width=18.0cm]{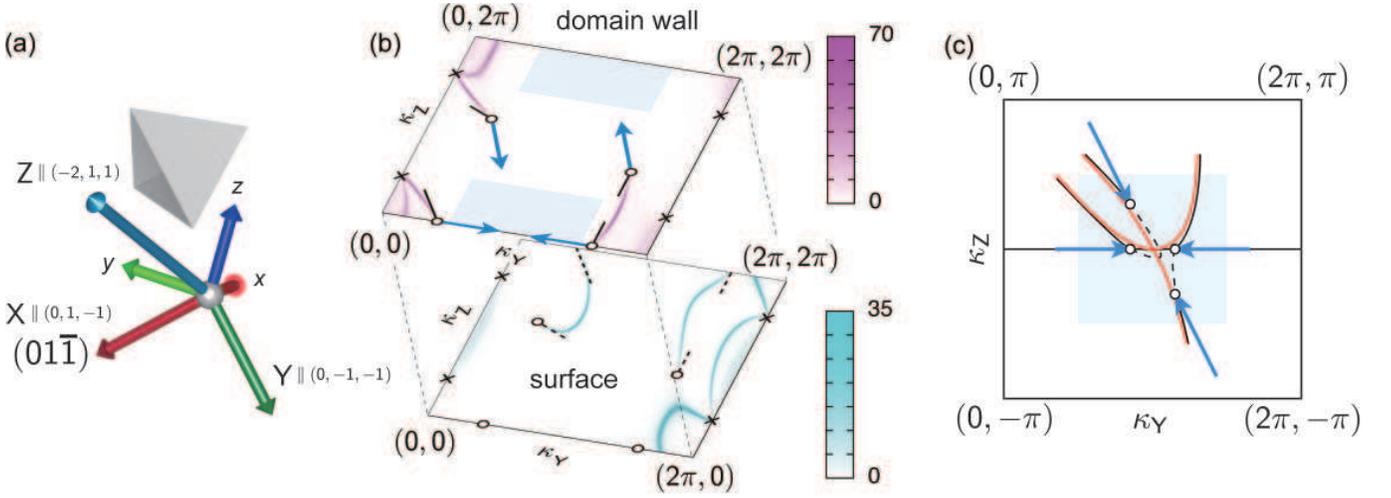}
\caption{
Summary of solutions for effective one-dimensional Dirac equations (2).
(a) Newly introduced coordinate axes $X$, $Y$, and $Z$ are illustrated for the $(01\overline{1})$-domain wall. 
(b) Zero modes of the chiral Dirac equations with low-energy $\vec{k}\cdot\vec{p}$-hamiltonians
including $\hat{h}^{(\pm)}_{\Gamma \kappa_ 0/\sqrt{3} (1,1,1)}$, Eq. (4) and $\hat{h}^{(\pm)}_{\Gamma \kappa_ 0/\sqrt{3} (-1,1,1)}$
(see Appendix \ref{SN3}).
Open circles indicate the 4 Weyl points projected to the $(\kappa_Y,\kappa_Z)$-plane
at
$\vec{k}_{\rm Weyl}=\pm\sqrt{|m|/2t}(1,1,1)$
and
$\vec{k}_{\rm Weyl}=\pm\sqrt{|m|/2t}(-1,1,1)$.
Crosses indicate
the
other 4 bulk Weyl points. 
Solid (broken) lines represent
the initial slopes of the loci of the
domain-wall (surface) states starting from the Weyl points obtained from Eq. (2),
where the full solution of
Eq. (1) is illustrated by color contour plot.
Small deviations of the
black solid (broken) lines from the expectation from the contour plot
may be ascribed to the small error arising from the reduction from the four-component Luttinger hamiltonian
to the two-component hamiltonian (2), where off-diagonal elements in the order of $k^2$ are ignored.
For $m\rightarrow 0$, solid curves shrink and disappear at $(0,0)$.
Arrows indicate the directions along which the projected Weyl points
move when
$|m|$ increases.
(c) Qualitative description for pair annihilation of bulk Weyl electrons on  $(\kappa_Y, \kappa_Z)$-plane.
Solid (broken) curves illustrate the loci of the domain  wall (surface) zero modes.
When the projected Weyl points move along the direction indicated by the arrows originally starting from $(0,0)$,
the pair annihilation occurs at $(\kappa_Y,\kappa_Z)=(\pi,0)$.
Then the closed loop of the Fermi surface on the domain walls appear,
which is represented by the red solid curves.
Note that the Brillouin zone is shifted $(0,\pi)$ from (b).
The shaded regions in (b) and (c) represent the same area.
\label{fig3}}
\end{figure*}

\section{ Weyl electrons and \red{their annihilation in pair.}}
\label{Section3}
Once the all-in/all-out orders are formed,
low-energy physics of the hamiltonian (1) with $\zeta < 0$,
therefore, physics of Weyl electrons, is essentially
captured by mean-field decouplings of the short-ranged Coulomb repulsion $U$
except quantitative corrections arising from gapped quantum and/or thermal spin fluctuations
and irrelevant quasiparticle renormalizations.

If the Coulomb repulsion $U$ is neglected,
low-energy electronic energy-momentum dispersion is described by
a variation of the Luttinger hamiltonian\cite{Luttinger56,Murakami04,Moon12},
which is a prototypical
four-component
effective
hamiltonian for semiconductors with the cubic symmetry.
The explicit form of the effective hamiltonian
directly derived from Eq.(1)
is given as
\eqsa{
 \hat{h}_{4\times 4} (\vec{k})
 =
 \left[+2t\left(1-\frac{k^2}{3}\right)-2\sqrt{2}|\zeta|\right]
 \mbox{\boldmath$1$}_4
 -2t
 \vec{d}(\vec{k})\cdot
 \vec{\hat{\Gamma}},
}
where
\eqsa{
 \vec{d}(\vec{k})^{T}=
 -
 \left(
 \frac{k_y k_z}{\sqrt{3}},
 \frac{k_z k_x}{\sqrt{3}},
 \frac{k_x k_y}{\sqrt{3}},
 \frac{k_x^2 -k_y^2}{2\sqrt{3}},
 \frac{
  3k_z^2 -k^2}{6}
 \right),\nn
}
and a vector of $4\times4$ Dirac matrices
$
 \vec{\hat{\Gamma}}^{T}=
 \left(
 \hat{\Gamma}^1 ,
 \hat{\Gamma}^2 ,
 \hat{\Gamma}^3 ,
 \hat{\Gamma}^4 ,
 \hat{\Gamma}^5
 \right)
$
defined as
\eqsa{
  \hat{\Gamma}^1
  &=&
  \left[
  \begin{array}{cc}
  0 & -i\hat{\sigma}_0 \\
  +i\hat{\sigma}_0 & 0 \\ 
  \end{array}
  \right],\nn
  \hat{\Gamma}^2
  &=&
  \left[
  \begin{array}{cc}
  0 & +\hat{\sigma}_z \\
  +\hat{\sigma}_z & 0 \\ 
  \end{array}
  \right],\nn
  \hat{\Gamma}^3
  &=&
  \left[
  \begin{array}{cc}
  0 & +\hat{\sigma}_y \\
  +\hat{\sigma}_y & 0 \\ 
  \end{array}
  \right],\nn
  \hat{\Gamma}^4
  &=&
  \left[
  \begin{array}{cc}
  0 & +\hat{\sigma}_x \\
  +\hat{\sigma}_x & 0 \\ 
  \end{array}
  \right],\nn
  \hat{\Gamma}^5
  &=&
  \left[
  \begin{array}{cc}
  +\hat{\sigma}_0 & 0 \\
  0 &-\hat{\sigma}_0 \\ 
  \end{array}
  \right]\nonumber
}
(see Appendix \ref{SN1} for derivation).
The low-energy bands are degenerate quadruply
at the crystallographic $\Gamma$-point, the center of the Brillouin zone in the momentum space
$\vec{k}=(0,0,0)$,
and form a so-called quadratic band crossing.

By adding a small but finite mean field term
\textcolor{black}{representing the all-in/all-out orders},
$m\hat{\Gamma}^{54}$, with another Dirac matrix $\hat{\Gamma}^{54}
=\left[\hat{\Gamma}^5 , \hat{\Gamma}^4 \right]/2i$ \textcolor{black}{(see Appendix A
for the microscopic derivation)}
and $m=Um_{\rm all}/2$,
eight
Weyl points at the momenta
$\vec{k}=\vec{k}_{\rm Weyl}\simeq \sqrt{|m|/2t}(\pm 1, \pm 1, \pm 1)$, up to order of $|m|$,
are induced instead,
while the four-fold degeneracy at the $\Gamma$-point
is lifted.
\textcolor{black}{Here,
$m_{\rm all}$ represents the amplitude of the magnetic moment at each site in the all-in/all-out phase.}
The energy spectrum is given through the poles of the Green's function,
\eqsa{
  \hat{G}_{4\times 4}(\vec{k},\omega)=
  \left[(\omega+\mu)\mbox{\boldmath$1$}_4-\hat{h}_{4\times 4}(\vec{k})\right]^{-1},
}
as
\eqsa{
&&  E(\vec{k})=-\mu
\nn
&& \pm\sqrt{4t^2 |\vec{d}(\vec{k})|^2 + m^2 \pm 4|m|t\sqrt{d_1(\vec{k})^2 + d_2(\vec{k})^2 + d_3(\vec{k})^2}},\nn
}
where $\mu$ is the chemical potential. The momenta of the Weyl points are given by the equations,
$d_4(\vec{k})=d_5(\vec{k})=0$ and $2t|\vec{d}(\vec{k})|=|m|$ (see Appendix \ref{SN2} for more details).
When the order parameter $m$ increases and becomes comparable to $t$,
these eight Weyl points
come closer
and are annihilated in pair at the
crystallographic L-points,
$\vec{k}=\vec{k}_{\rm L}=(\pm \pi/4a, \pm \pi/4a, \pm\pi/4a)$,
at the boundary of the Brillouin zone. We note that
there are only four inequivalent L-points.

To understand nature of bulk Weyl electrons, their pair-annihilation,
Fermi arcs on surfaces, and ones on domain walls, it is sufficient
to employ a $\vec{k}\cdot\vec{p}$-perturbation theory,
which is a traditional technique for semiconductor and its interface physics\cite{Dresselhaus}
around the Weyl points,
starting form the four-component effective hamiltonian, $\hat{h}_{4\times 4}(\vec{k})+m\hat{\Gamma}^{54}$. 
Especially, nearby the $\Gamma$-point and the L-points, the $\vec{k}\cdot\vec{p}$-perturbation theory
gives us simple expressions suitable for exploration of physics of interfaces,
as shown in Sec.\ref{Section4}.
As unperturbed wave functions
for the $\vec{k}\cdot\vec{p}$-perturbation theory, we choose the
wave functions at these symmetric points classified by irreducible representations
of the point groups T$_{\rm d}$ and D$_{\rm 3d}$ for the $\Gamma$-point and the L-points, respectively.
Around the $\Gamma$-point, the $\vec{k}\cdot\vec{p}$-hamiltonian expanded from
the quartet labeled by
G$_{3/2}$ in terminology of the point group T$_{\rm d}$
is nothing but the Luttinger hamiltonian discussed above.
The four-component Hilbert space labeled by
irreducible representations E$_{\rm 3/2u}$ $\oplus$E$_{\rm 1/2g}$ at the L-point
gives the basis for the $\vec{k}\cdot\vec{p}$-hamiltonian
around the L-points. 
Once the $\vec{k}\cdot\vec{p}$-perturbation theory around the Weyl points nearby the
$\Gamma$-point and L-points is obtained,
it leads to the entire description of the Weyl electrons through
the interpolation of these low-energy theories around the $\Gamma$-point and the L-points.

Here we elucidate the relevance of our model hosting the 8 Weyl points
to the pyrochlore iridium oxides $R_2$Ir$_2$O$_7$.
First of all, the quadratic band crossing and
four-fold degeneracy at the $\Gamma$-point are symmetry-protected properties
of the $J_{\rm eff}$=1/2-manifold on the pyrochlore lattice.
When the time-reversal symmetry is broken with keeping the $T_{\rm d}$-symmetry,
the 8 Weyl points immediately stem from the $\Gamma$-point.
When the system becomes insulating,
the pair-annihilation of the 8 Weyl points necessarily occur
through the level crossing at the L-points
between the Zeeman-splitted states from E$_{\rm 1/2g}$- and E$_{\rm 3/2u}$-doublets
discussed in the paragraph above.

On the other hand, 24 Weyl points
are found by the LSDA+SO+$U$ calculation in Ref.\onlinecite{Wan11}.
The 24 Weyl points are created as three pairs at each of four L-points when
the splitted states from E$_{\rm 1/2g}$- and E$_{\rm 1/2u}$-doublets at the L-points show level crossings,
instead of states from E$_{\rm 1/2g}$- and E$_{\rm 3/2u}$-doublets.
These 24 Weyl points are nothing to do
with the 8 Weyl points discussed in this paper. 
As detailed in the following section,
what we found
is that the 8 Weyl points leave gapless domain-wall states as
a topological nature of pyrochlore iridium oxides even after their pair-annihilation
while the 24 Weyl points leave nothing.
Furthermore, whether the crossing of the E$_{\rm 1/2g}$- and E$_{\rm 1/2u}$-doublets occurs
depends on details of material  parameters and does not always happen.
The 8 Weyl points are not focused on in Ref.\onlinecite{Wan11} 
although signs of the 8 Weyl
points are found inside the semimetal phase
with Fermi surfaces in the band dispersion.

\section{ \red{Effective one-dimensional Dirac hamiltonian}}
\label{Section4}
Under the influence of a small but finite order parameter $m$,
the four-component effective hamiltonian, $\hat{h}_{4\times 4}(\vec{k})+m\hat{\Gamma}^{54}$,
exhibit
Weyl points
consisting of two of the four components
while the other
two components are gapped
(see Eq.(5) and Appendix \ref{SN3}).
Around the Weyl points, the hamiltonian \red{therefore} breaks
up into a pair of two-component hamiltonians, \tmag{up to the linear-order in the $\vec{k}\cdot\vec{p}$-perturbation},
one of which is nothing but a Weyl hamiltonian describing three-dimensional massless fermions.
For $m>0$, we note the Weyl hamiltonian as $\hat{h}_{\Gamma\vec{k}_{\rm Weyl}}^{(+)}$ ($\hat{h}_{{\rm L}\vec{k}}^{(+)}$),
and describe the other two-component \red{gapped part} as $\hat{h}_{\Gamma\vec{k}_{\rm Weyl}}^{(-)}$ ($\hat{h}_{{\rm L}\vec{k}}^{(-)}$),
around the $\Gamma$-point (the L-points).
For instance, the four-component effective hamiltonian
is expanded with \tmag{respect to} the momentum measured from the Weyl points $\delta\vec{k}$ as 
\eqsa{
&\hat{h}_{4\times 4}(\vec{k}_{\rm Weyl}+\delta\vec{k})+m\hat{\Gamma}^{54}
 \rightarrow
\nn
&\left[
\begin{array}{cc}
\hat{h}_{\Gamma\vec{k}_{\rm Weyl}}^{(+)}(\delta\vec{k}) & \mathcal{O}(t |\delta\vec{k}|) \\
 \mathcal{O}(t |\delta\vec{k}|) & \hat{h}_{\Gamma\vec{k}_{\rm Weyl}}^{(-)}(\delta\vec{k})\\
\end{array}
\right]
+\mathcal{O}(t^2 |\delta\vec{k}|^2),
}
around the $\Gamma$-point, \tmag{after an appropriate unitary transformation independent of $m$}.
Since the gapped part $\hat{h}_{\Gamma\vec{k}_{\rm Weyl}}^{(-)}(\delta\vec{k})$ at $\delta\vec{k}=\vec{0}$ has
two eigenvalues $\pm 2m$ (see Eq.(5)) \tmag{with $m>0$,
the low-energy excitations up to $t|\delta\vec{k}|$ is described by the Weyl hamiltonian $\hat{h}_{\Gamma\vec{k}_{\rm Weyl}}^{(+)}$
at a small $\delta\vec{k}\neq\vec{0}$.} 
Surprisingly, when we change the sign of $m$ with keeping its amplitude, or simply apply the time reversal
operator to the hamiltonian,
the role interchanges and the two-component hamiltonian $\hat{h}_{\Gamma\vec{k}_{\rm Weyl}}^{(-)}$ ($\hat{h}_{{\rm L}\vec{k}}^{(-)}$) describes
Weyl electrons while $\hat{h}_{\Gamma\vec{k}_{\rm Weyl}}^{(+)}$ ($\hat{h}_{{\rm L}\vec{k}}^{(+)}$) describes gapped
components.

As elucidated in the literature\cite{Wan11,Silaev12},
bulk Weyl electrons result in Fermi arcs on surfaces and/or domain walls of the bulk crystals.
By using the $\vec{k}\cdot\vec{p}$-hamiltonians
$\hat{h}_{\Gamma\vec{k}_{\rm Weyl}}^{(\pm)}$ or $\hat{h}_{{\rm L}\vec{k}}^{(\pm)}$ around
the Weyl points,
we can sketch the Fermi arcs not only on the surfaces
but also on the magnetic domain walls in the following\textcolor{black}{.} 
\begin{figure*}[ht]
\centering
\includegraphics[width=17.5cm]{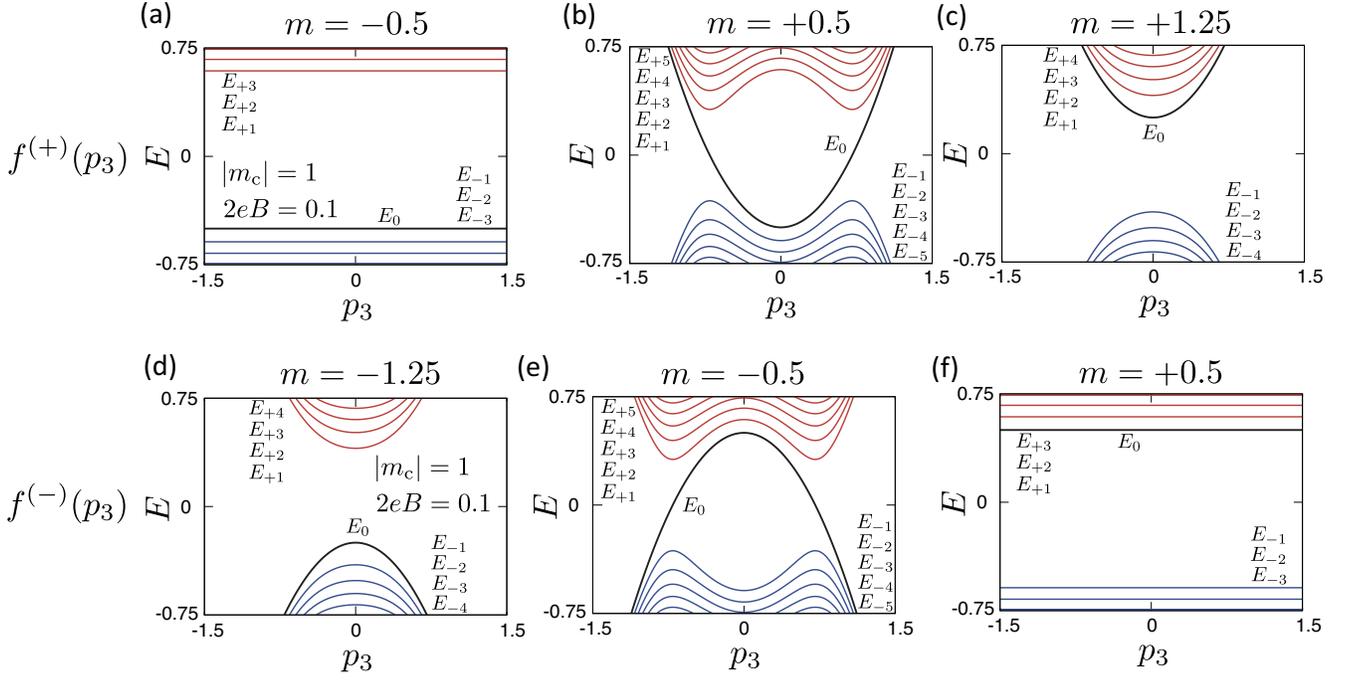}
\caption{
Landau level spectra for the 3D Weyl equation Eq.(11).
Throughout this Figure, we choose $|m_{\rm c}|=1$ and $2eB=0.1$.
The Landau levels with $f^{(+)}(p_3)$ are shown for $m=-0.5$ (a), $m=+0.5 < |m_{\rm c}|$ (b),
and $m=1.25 > |m_{\rm c}|$ (c).
The Landau levels with $f^{(-)}(p_3)$ are also illustrated
for $m=-1.25 < -|m_{\rm c}|$ (d), $m=-0.5 > -|m_{\rm c}|$ (e),
and $m=+0.5$ (f).  
The Landau levels $E_n$ with $n>0$ ($n<0$) are represented by red (blue) solid lines.
The 0-th Landau levels $E_0$ are shown by black solid lines.
For $m=\pm 0.5$, there are remnants of the Weyl nodes around $E=0$:
The 0-th Landau levels traverse $E=0$ from the valence bands to the conduction bands.
The asymmetric nature of the 0-th Landau levels around $E=0$ is a manifestation
of the chiral anomaly.
\label{figW}}
\end{figure*}
\begin{figure*}[htb]
\centering
\includegraphics[width=18.0cm]{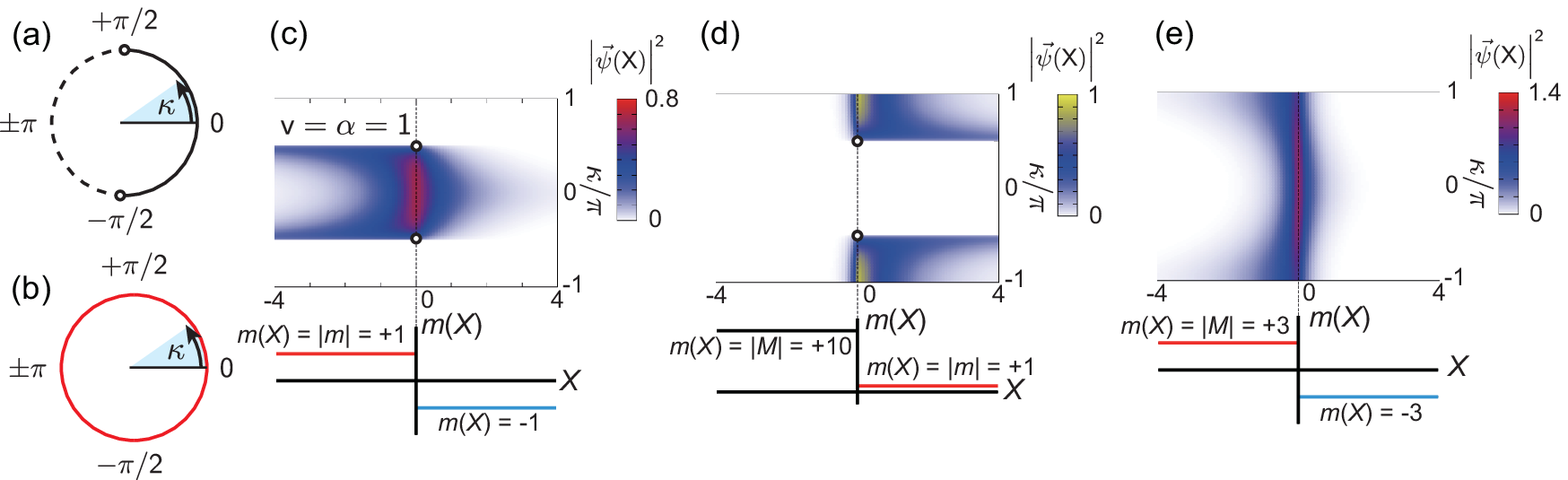}
\caption{
Solutions for effective one-dimensional chiral Dirac equations (4).
(a) and (b) Simplified degrees of freedom $\kappa$
corresponding to one parameter representation of loci of zero modes on  $(\kappa_Y,\kappa_Z)$.
Solid (broken) curve illustrates the loci of the domain-wall (surface) zero modes.
White circles indicate the Weyl points.
(a) represents the case before the pair-annihilation of the Weyl points,
corresponding to (c) and (d).
On the other hand, (b) after the pair-annihilation
corresponds to (e),
where the surface loci is eaten up by the domain-wall loci and the domain-wall loci form a closed Fermi surface in the Brillouin zone. 
(c)
Wavefunction amplitude for a domain wall state for the chiral Dirac equation (4)
with $\alpha_{+}$ in the plane of the real space coordinate $X$ and momentum coordinate $\kappa$, with
$v=\alpha_{+}=1$ and $m(x)=- \theta (+X)+\theta (-X)$.
The ``arc" state at the domain wall penetrates into one side of the bulk at the Weyl points.
Another degenerate ``arc" state localized in the other nearby side of the domain wall,
obtained with $v=1$ and $\alpha_{-}=-1$, penetrates to the other side of the bulk (not shown).
(d) Wavefunction amplitude for a surface state for the chiral Dirac equation (4), with
$v=\alpha=1$ and $m(x)=\theta (+X)+10\theta (-X)$. The ``arc" is formed in the missing part of the domain-wall arc.
(e) Wavefunction amplitude for a domain wall state for the chiral Dirac equation (4), with
$v=\alpha=1$ and $m(x)=-3\theta (+X)+3\theta (-X)$.
The domain wall state now forms the closed loop of the Fermi surface without penetration into the bulk. 
For $|m(X)|>2$, there are no zero modes for surfaces.
\label{fig5}}
\end{figure*}

For \red{clarification}, we concentrate on a pair of Weyl points,
$\vec{k}_{\rm Weyl}=\pm \sqrt{|m|/2t} (1,1,1)$ with $|m|/t\ll 1$,
and on a surface or domain wall perpendicular to $(0,+1,-1)$,
namely, $(01\overline{1})$-surface or domain wall.
In the following discussion, we call a coordination axis along $(0,+1,-1)$
as $X$-axis
and introduce an oblique coordinate $(X,Y,Z)$ together with
the corresponding reciprocal momentum coordinate $(\kappa_X,\kappa_Y,\kappa_Z)$
defined through
\eqsa{
 \vec{r}=X
 \left[
 \begin{array}{c}
  0\\
  +2a\\
  -2a\\
  \end{array}
  \right]
  + Y
 \left[
 \begin{array}{c}
  0\\
  -2a\\
  -2a\\
  \end{array}
  \right]
  +Z
 \left[
 \begin{array}{c}
  -4a\\
  +2a\\
  +2a\\
  \end{array}
  \right],
}
and
\eqsa{
 \vec{k}=\kappa_X
 \left[
 \begin{array}{c}
   0\\
   +1/4a\\
   -1/4a\\
  \end{array}
  \right]
  + \kappa_Y
 \left[
 \begin{array}{c}
  -1/4a\\
  -1/4a\\
  -1/4a\\
  \end{array}
  \right]
   +\kappa_Z
 \left[
 \begin{array}{c}
  -1/4a\\
   0\\
   0\\
  \end{array}
  \right]\nn
}
(see also Fig.~\ref{fig3}(a) and Appendix \ref{ADWs}).
As detailed above, around
the Weyl points,
the
Luttinger hamiltonian
breaks up into a pair of
the following
two component Dirac hamiltonians
$\hat{h}_{\Gamma \vec{k}_{\rm Weyl}}^{(+)}$ and $\hat{h}_{\Gamma \vec{k}_{\rm Weyl}}^{(-)}$ that
describe low-energy physics in the all-out and all-in domain,
with $m > 0$ and $m < 0$, respectively.
(See Appendix \ref{SN1})
For $\vec{k}_{\rm Weyl}= \kappa_0 (1/\sqrt{3},1/\sqrt{3},1/\sqrt{3})$
with $\kappa_0=\pm\sqrt{3|m|/2t}$,
the
two-componet
Dirac hamiltonian
up to the linear order in $-i\partial_X$, $\delta\kappa_Y$, and $\kappa_Z$,
is given as
\eqsa{
 &&\hat{h}_{\Gamma\vec{k}_{\rm Weyl}}^{(\pm)}(-i\partial_X, \delta\kappa_Y, \kappa_Z ; X)
 =
 h_0 (\delta\kappa_Y, \kappa_Z)\hat{\sigma}_0
 \nn
 &&
 +
 h_x (\delta\kappa_Y, \kappa_Z)\hat{\sigma}_x
 +
 h_y (-i\partial_X) \hat{\sigma}_y
 +
 h_z^{(\pm)} (\delta \kappa_Y,\kappa_Z, X) \hat{\sigma}_z,\nn
\label{DiracGamma}
}
where we
introduced a new variable $\delta \kappa_{Y}$
defined through
$(\kappa_X, \kappa_Y=\kappa_0/\sqrt{3}+\delta\kappa_Y,\kappa_Z)$
and replaced $\kappa_X$ with $-i\partial_X$ (see Appendix \ref{SN3} for derivation).
The Weyl points are projected to $(\delta\kappa_Y , \kappa_Z)=(0,0)$.
Here coefficients of the identity matrix
and Pauli matrices in the Dirac hamiltonian (\ref{DiracGamma})
are derived from the original hamiltonian (1), via the low-energy Luttinger hamiltonian, as
$h_0 = -4\sqrt{3}t\kappa_0 (\delta\kappa_Y+\kappa_Z/3)$,
$h_x = 4\sqrt{3}t\kappa_0 \kappa_Z/3$, $h_y=4t\kappa_0 i \partial_X$,
and
$h_z^{(\pm)} = \mp 4t\kappa_0 (\delta \kappa_Y + \kappa_Z /3 )/\sqrt{3}+m(X)\mp |m|$.

Then the two component one-dimensional
Dirac equation,
$\hat{h}^{(+)}_{\Gamma\vec{k}_{\rm Weyl}}
\vec{\psi}(X)
 =E
\vec{\psi}(X)
$
gives description of
bound states on the surface or domain walls
by introducing suitable $X$-dependent ``mass" terms $m(X)$\cite{JackiwRebbi76,FanZhang}.
Here, 
the all-out (all-in) domain is described by $m(X)=+|m|$ ($m(X)=-|m|$).
We also note
that, if $|m|$ is large enough,
the Weyl points are annihilated in pair and the bulk system becomes a trivial
magnetic insulator.
Therefore, 
the mass term, $m(X)=
  |m| \theta ( - X )
  -|m| \theta ( X) $, gives a description of magnetic domain wall at $X=0$
for long-wave-length behaviors.

The $X$-dependent mass term for the magnetic domain walls
introduced above indeed reproduces the numerical solution of the tight-binding hamiltonian Eq.(1) for
the Fermi arcs (solid curves) around the Weyl points (white circles)
\tmag{projected} onto the domain-wall Brillouin zone, at least up to linear order, as shown in Fig.\ref{fig3}(b).
It shows the validity of the effective one-dimensional Dirac equation for the domain-wall Fermi arcs.

Then, we explain
how a description for 
a surface between a vacuum ($X<0$)
and the bulk ($X>0$)
can be mimicked by
$
  m(X)=
  M\cdot{\rm sign} (m)\theta ( -X)
  +m \theta ( X)$ with $M\gg |m|$.
\textcolor{black}{The introduction of the large amplitude of the mass $M$ without the sign change in $m(X)$
mimics a zero Fermi velocity limit and indeed offers
an effective description of vacuum.
In the 1D Chern insulator, topologically trivial phases with the zero Chern number
are realized by setting the Fermi velocity equal to zero.
By taking into account the fact that the relevant length scale
governing the wave functions of the edge states
is the ratio of the amplitude of the mass and the Fermi velocity,
the small Fermi velocity limit corresponds to the large mass amplitude limit
independently of the sign of the mass $M$. 
}
The comparison with the numerical solution of the tight-binding hamiltonian Eq.(1)
at the surface between the bulk and the real vacuum 
\textcolor{black}{indeed} 
supports the validity of the mass term $m(X)$, as shown in Fig.\ref{fig3}(b):
The Fermi arcs (broken curves) obtained with the mass term $m(X)$
around the projected Weyl points (white circles)
are consistent with the numerical solution of Eq.(1), up to linear order.

When we treat three successive boundaries such as ones between a vacuum and an all-out domain,
an all-out domain and an all-in domain, and between an all-in domain and a vacuum, we encounter a superficial problem
with the choice of the mass term $m(X)\hat{\sigma}_z$ introduced above for a boundary between a vacuum and bulk:
The vacua at the both ends of the system have a different sign of $M$, and therefore, are seemingly not
connected each other. In other word, there seem to exist two different vacua, which seems to be unphysical.
However, if we take a atomic limit only inside the vacua described by the mass term,
we can rotate the mass term and can connect these two vacua by
applying unitary transformations such as $e^{i\pi \hat{\sigma}_x /2}$ and changing the sign of $M$. 
We also remind the readers that the edge-state wave functions change smoothly when the atomic limit are taken
(see concrete examples of the wave functions such as Eq.(D28)), while the vacua described by the
mass term do not show the chiral anomaly\cite{Nielsen}
at the atomic limit.

As the order parameter $m$ develops, the two Weyl points
at $\vec{k}=+|\kappa_0|(1/\sqrt{3},1/\sqrt{3},1/\sqrt{3})$ and
$\vec{k}=-|\kappa_0|(1/\sqrt{3},1/\sqrt{3},1/\sqrt{3})$ come closer,
and, finally, are annihilated in pair at a L-point $\vec{k}_{\rm L}=(\pi/4a,\pi/4a,\pi/4a)$.
Around the L-point, the pair of the two-component Dirac hamiltonian
is given as
\eqsa{
  \hat{h}^{(\pm)}_{{\rm L}\vec{k}_{\rm L}}=
   h_x (\kappa_Z) \hat{\sigma}_x
   +
   h_y (-i\partial_X, \kappa_Z) \hat{\sigma}_y
   +
   h_z^{(\pm)} (X) \hat{\sigma}_z,
   \label{ChiralDiracL}
}
where the coefficients of the Pauli matrices $h_x$, $h_y$ are linear functions
of their arguments, and $h_z^{(\pm)} = -m(X)\pm |m|$.
\textcolor{black}{T}he above Dirac hamiltonians (\ref{ChiralDiracL})
do not 
\textcolor{black}{contain linear terms of $\delta\kappa_Y$,}
where
$(\kappa_Y,\kappa_Z)=(\pi+\delta\kappa_Y,\kappa_Z)$, and, thus,
the pair-annihilation point is given by $(\delta\kappa_Y,\kappa_Z)=(0,0)$.
Therefore, with a condition $\kappa_Z=0$ or $h_x =0$,
the Dirac hamiltonian (\ref{ChiralDiracL}) possesses chiral symmetry
with a chiral operator $\hat{\sigma}_x$
(see Appendix \ref{SN4} for derivation and topological
properties of the Dirac equation).

Figure \ref{fig3}(b) illustrates an example how
the domain-wall (solid curves) and surface (broken curves)
states extend around the Weyl points (white circles)
before the pair-annihilation at $(\kappa_Y, \kappa_Z)=(\pi,0)$ shown in Fig.\ref{fig3}(c).

\section{Chiral anomaly}
\label{newV}
In addition to the 2-component 1D Dirac equations described above,
the quantum chiral anomaly originating from the bulk Weyl nodes\cite{Nielsen}
also confirms
the emergence of the domain-wall states.
Below, we explain that the chiral anomaly due to bulk Weyl nodes
leaves their trace even after their pair-annihilation, which
inevitably induces domain-wall states.

Following Nielsen and Ninomiya (Ref.\onlinecite{Nielsen}), we start with Weyl fermions
coupled to an external magnetic filed. 
Pair-wise annihilation of two Weyl nodes with opposite chiralities
coupled to an external magnetic field $(0,0,B)$ (or a vector potential
$(0,Br_1,0)$)
is modeled by the following three-dimensional Weyl equation
\eqsa{
\left[
-i\partial_{r_1}\hat{\sigma}_x + (p_2 -eBr_1)\hat{\sigma}_y \textcolor{black}{+} f(p_3)\hat{\sigma}_z
\right]
\psi
=
E
\psi
}
where
we introduce a real-space Cartesian coordinate $(r_1, r_2, r_3)$ and corresponding momentum
coordinate $(p_1,p_2,p_3)$, which are connected through $p_a\leftrightarrow -i\partial_{r_a}$ ($a=1,2,3$).
The microscopic origin of the above 3D Weyl equation is explained in Appendix E
(see Eq.(\ref{H44LL}) and following paragraphs)
although the 3D Weyl equation is a general one that describes the pairwise annihilation
of Weyl nodes.
The function depending on the third momentum coordinate $f(p_3)$ determines
the chirality of the Weyl nodes and includes a mass term controlling the pairwise annihilation.
For example, if we concentrate on the pair-annihilation at a L-point $(\pi/4a, \pi/4a, \pi/4a)$
in the all-out domain with $m>0$,
we can choose the function as \textcolor{black}{$f(p_3)=+p_3^2 - |m_{\rm c}|+m$}, where $|m_{\rm c}|$ is
a critical amplitude of
the all-in/all-out magnetic moment for the pair-annihilation.
Here, the pair-annihilation point, namely the L-point is represented by $p_3=0$ in the
newly introduced momentum coordinate.
The above Weyl equation leads \tmag{to} the following set of eigenvalues describing the Landau levels,
\textcolor{black}{$E_0={\rm sign}(B)f(p_3)$} and
$E_n={\rm sign}(n)\sqrt{f(p_3)^2 + 2e|B| |n|}$,
where $n$ is non-zero integer as $n=\pm 1, \pm 2, \pm 3, \dots$.
The emergence of the Landau level $E_0$ is nothing but a manifestation of
the chiral anomaly. 

Then let us go into a detailed description on the domain wall
based on the bulk Weyl equation
introduced above.
As already discussed, the low-energy physics of the all-in/all-out
phases are described by a pair of 2-component Dirac equations:
For a description on domain walls, we introduced
the 1D Dirac hamiltonians $\hat{h}^{(\pm)}_{{\rm L}\vec{k}_{\rm L}}$.
Inside an all-out domain with $0<m(<|m_{\rm c}|)$,
$\hat{h}^{(+)}_{{\rm L}\vec{k}_{\rm L}}$
describes gapless excitations while $\hat{h}^{(-)}_{{\rm L}\vec{k}_{\rm L}}$
describes gapped excitations. On the other hand, inside an all-in domain
with $0>m(>-|m_{\rm c}|)$, $\hat{h}^{(-)}_{{\rm L}\vec{k}_{\rm L}}$
corresponds to gapless ones while $\hat{h}^{(+)}_{{\rm L}\vec{k}_{\rm L}}$ 
corresponds to gapped ones.
\textcolor{black}{(For its illustration, see later discussion at Fig.~\ref{fig9}.)}
For replacing the 1D Dirac equations by bulk 3D Weyl equation,
we introduce a set of $f(p_3)$'s as
\textcolor{black}{
\eqsa{
f^{(+)}(p_3)&=&\theta (m)\left[+p_3^2 -|m_{\rm c}|+m\right]\nn
&&+\theta (-m)\left[-|m_{\rm c}|+m\right],
}
}
and
\textcolor{black}{
\eqsa{
f^{(-)}(p_3)&=&\theta (m)\left[+|m_{\rm c}|+m\right]\nn
&&+\theta (-m)\left[- p_3^2 +|m_{\rm c}|+m\right].
}
}
Then, we obtain Landau levels illustrated in Fig.\ref{figW} with typical parameter sets.
Here, we note that, even after the pair-annihilation with $|m|>|m_{\rm c}|$,
the 0-th Landau level $E_0$ remains asymmetric while other Landau levels $E_n$ are symmetric
around the zero energy. 

The structure of the Landau levels given by using Eqs.(11), (12), and (13), illustrated in Fig.\ref{figW},
directly leads to the emergence of the gapless domain-wall states.
From the Landau level spectrum in Fig.\ref{figW}, the 0-th Landau levels for both of $f^{(+)}(p_3)$ and $f^{(-)}(p_3)$
appear \textcolor{black}{above} $E=0$ and
\textcolor{black}{at bottom of the conduction bands}
in the all-out domain with $m>0$.
On the other hand, these 0-th Landau levels appear
\textcolor{black}{below}
$E=0$ and
\textcolor{black}{on top of the valence bands} 
in the all-in domain with $m<0$.
As a result, if the all-in and all-out domain are smoothly connected each other,
these 0-th Landau levels are also smoothly connected and result in two gapless domain-wall states. 
Here we remind the readers important facts that the two-dimensional Hilbert subspaces
described by the 3D Weyl equations with $f^{(\pm)}(p_3)$ are orthogonal each other, and
the eigenvectors of the eigenvalues $E=E_0$ are orthogonal to those of the other Landau levels
independently of the choice for $p_3$, $m$, and $eB$.
\textcolor{black}{We further detail in Sec.VIII the symmetry protection of the orthogonality
by using the explicit symmetry satisfied by the domain walls.}

Therefore, in addition to the topological nature of the 1D Dirac equations,
the chiral anomaly as a bulk property protects the emergence of the gapless domain-wall states
even after the pair-annihilation of the bulk Weyl nodes.
In both descriptions on the emergence of the domain-wall states, the domain-wall states
are proven to appear in pair.
In contrast to the surface states of strong topological insulators,
the degenerated or pair-wise domain-wall states allow occurrence of the Anderson localizations
\textcolor{black}{by the impurities that breaks the translational symmetry,}
and degeneracy liftings due to additional spontaneous symmetry breakings.
We note that similar domain-wall states are proposed in graphene with a broken inversion symmetry,
characterized by asymptotic valley-resolved Chern numbers\cite{Yao}, thus, by the parity anomaly.

\section{Topological properties of domain-wall states in a simplified model}
\label{Section5}
If the translational invariance along the domain walls is preserved,
\textcolor{black}{the}
ingap states
\textcolor{black}{at these domain walls}
are protected
by the chiral symmetry\cite{RyuSchnyder,Teo} of the Dirac hamiltonian
particularly at a pair-annihilation point \textcolor{black}{$(\kappa_Y, \kappa_Z)$=$(\pi, 0)$}
(see Appendix \ref{SN4} \textcolor{black}{
and also Sec.VIII for the protection by the symmetries}), and by a generalized chiral symmetry\cite{Kawarabayashi} at other $k$-points.
After the pair-annihilation, only the loci of the domain-walls survive.
\begin{figure*}[htb]
\centering
\includegraphics[width=18.0cm]{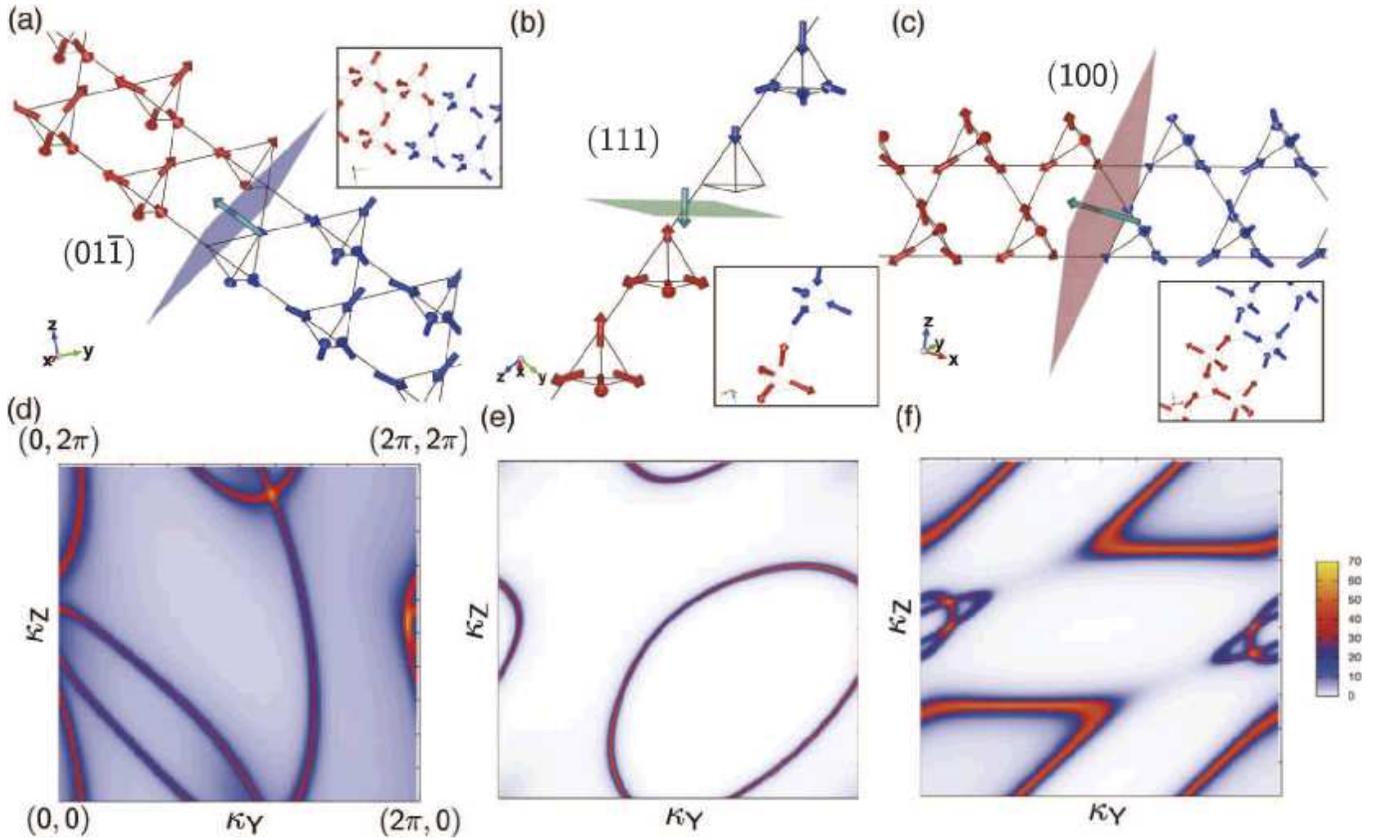}
\caption{
Magnetic domain walls and domain wall states.
(a)-(c)
Optimized magnetic structures for $(01\overline1)$, $(111)$, and $(100)$-domain walls,
obtained from initial configurations shown in the insets, for $U/t=4$, $\zeta/t=-0.2$,
and $k_{\rm B}T/t=0.1$.
The length and direction of the arrows are determined by optimized unrestricted
Hartree-Fock solutions.
We note that only a part of the supercells nearby the domain walls are illustrated.
Uniform magnetization $m_0$ induced
by the insertion of
the domain walls and the intersection
of the domain walls with the supercells are also illustrated as green bold arrows and shaded planes, respectively.
(d)-(f) Spectral functions projected to the domain walls for
(d) the $(01\overline{1})$-domain wall, (e) the $(111)$-domain wall,
and (f) the $(100)$-domain wall, for $k_{\rm B}T/t=0.1$, with a finite Lorentzian width $\delta=0.01t$.
At this temperature,
bulk Weyl points do not exist any more.
Every domain wall contributes to ingap states at the chemical potential or the Fermi level, forming open two-dimensional Fermi surfaces.
As a consequence of the pair-annihilation of the Weyl points shown in Fig.~\ref{fig3}(c),
the Fermi arcs for the domain walls are now closed at this temperature.
The complexity of the domain-wall Fermi surfaces
originates from the following fact:
Depending on the orientation of the domain walls,
there are many choices how to connect the
Weyl points with the loci of the domain-wall zero modes.
\label{fig6}}
\end{figure*}

\begin{figure*}[htb]
\centering
\includegraphics[width=18.0cm]{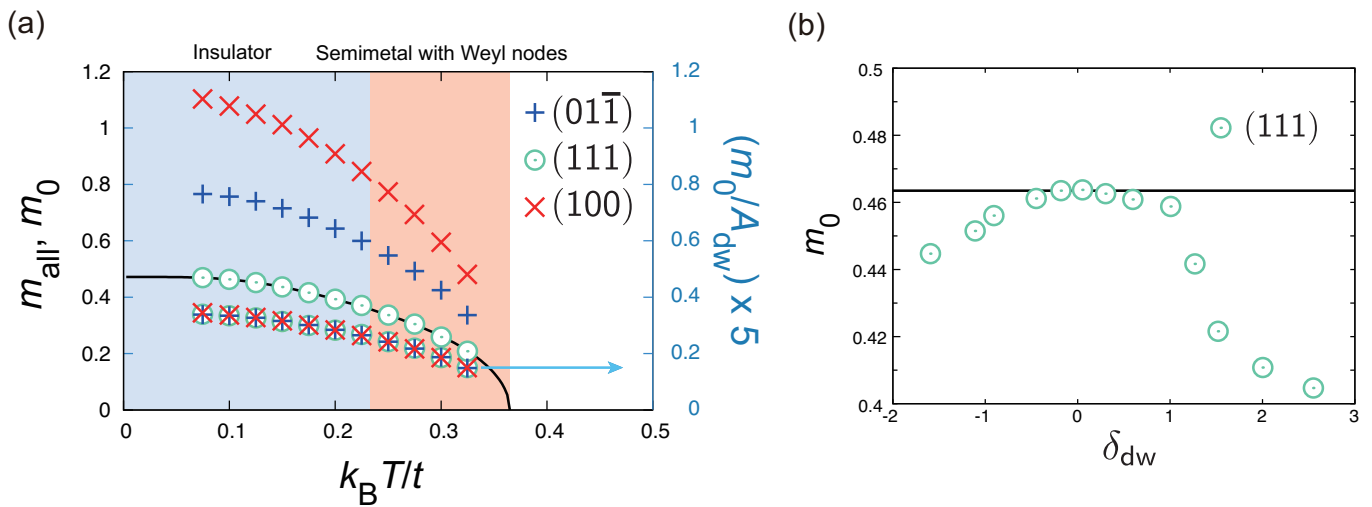}
\caption{
Uniform magnetization $m_0$ induced around the domain walls.
(a)
Uniform magnetization $m_0$ induced around the domain walls per
supercell
shown as top three symbols at each temperature for three different domain wall planes, 
in comparison with the bulk
all-in/all-out magnetic ordered moment $m_{\rm all}$ (solid black curve).
The uniform magnetization per unit area
($a^2$)
of the domain walls
$m_0/A_{\rm dw}$
is also shown as three lowest symbols at each temperature.
Here 
the area of the intersection of a supercell with the domain walls, $A_{\rm dw}$, 
for the $(01\overline{1})$, $(111)$, $(100)$-domain wall
are $8\sqrt{2}$, $4\sqrt{3}$, and $16$ in units of $a^2$, respectively.
These three data points are almost on top of each other.
(b) Doping dependence of uniform magnetization of $(111)$-domain wall for
$k_{B}T/t=0.1$.
The doping $\delta_{\rm dw}$ is defined as increase or decrease of the electron number
per present supercell
around the domain walls due to changes in the chemical potential.
It shows that the uniform magnetization is insensitive
to the charge doping around the domain wall, within $|\delta_{\rm dw}|<1$.
When more than one electron is doped in the supercell around the domain wall,
the reduction of the uniform magnetization becomes significant with an asymmetry between electron and hole dopings.
\label{fig7}}
\end{figure*}

The essential physics of these bound states
is captured by the following
toy model with higher symmetry, namely, a pair of chiral
Dirac equations that
describe 1D Chern insulators\cite{RyuSchnyder,Teo} defined by
\eqsa{
\left\{
\left[ 
\alpha_{\pm} (1-\cos \kappa)
-m(X)\right]\hat{\sigma}_z
 +vi\hat{\sigma}_y \partial_X
\right\}
\vec{\psi}(\vec{X})
 =
 E
\vec{\psi}(\vec{X}),
\label{ChD2}
}
where
$\alpha_{\pm}=\pm \alpha $ $(\alpha>0)$ and
$\kappa$ represents the degrees of freedom of $\kappa_Y$ and $\kappa_Z$.
The two ``Weyl" points appear at $\kappa$ that satisfies $\alpha (1-\cos \kappa)-m(X)=0$ (see Fig.~\ref{fig5}). 

When we approach the pair annihilation of Weyl points, the Fermi arcs
are expected to shrink on the surface. On the contrary, as confirmed later numerically,
the Fermi arcs on the domain walls becomes elongated by eating a part of the former arc on the surface.
Furthermore, after the pair annihilation, they form
a closed loop (open Fermi line connected through equivalent Brillouin zone boundaries).
As is clear in the high-symmetry model (\ref{ChD2}),
after the pair annihilation of the Weyl points,
the surface between a vacuum and the domain
is no longer a topological boundary:
For a given $\kappa$, boundaries where the mass changes are
classified in the terminology of the 1D Chern insulators.
The topological invariant changes its sign at the domain walls, as
the 1D \textcolor{black}{weak} Chern insulators
\textcolor{black}{(see also Sec.VIII for detailed symmetry analysis)}.

Here we note that, although there exists a substantial similarity of the present Dirac hamiltonian to
the well-studied Su-Schrieffer-Heeger hamiltonian (see Ref.\onlinecite{SSH}), namely, the chiral symmetry shared by both hamiltonian,
the latter effective hamiltonian for polyacetylene and other conducting polymers additionally possesses
the time-reversal and particle-hole symmetries.
In addition to the difference in the symmetric properties, our 1D chiral Dirac
equations describe \gr{the domain-wall states at a specified $\kappa$. By a variation of $\kappa$, they constitute 2D Fermi surfaces on the domain walls },
while the edges of the Su-Schrieffer-Heeger model are \gr{genuinely} zero dimensional ones.

\section{ Unrestricted Hartree-Fock analysis}
\label{Section6}
The prediction based on the simple
Dirac equations
is confirmed by using fully unrestricted Hartree-Fock
analysis
(see Appendix \ref{AUHF})
of the original hamiltonian (\ref{TB}) on the large supercell calculations with
three different and typical domain walls, namely,
$(01\overline{1})$, $(100)$, and $(111)$-domain walls,
with a typical parameter set, $U/t=4$ and $\zeta/t=-0.2$
(see Appendix \ref{ASCs} for definition of the supercells).
The self-consistent solution with optimized magnetic moment and
charge distribution retains gapless domain-wall states
in general and indeed on these three examples (see Fig.~\ref{fig6}).

Moreover, these domain walls
bring about
uniform magnetizations
perpendicular to themselves,
which are defined as the sum of magnetic moments within the supercells.
Surprisingly, the amplitude of these magnetizations per unit area of the domain wall
does not depend on the direction of the domain wall and only depends on $m$
within numerical errors (Fig.~\ref{fig7}(a)).
As a result, although
insertion of a single domain wall brings about a uniform nonzero magnetization,
the total magnetization of a closed domain wall surrounding a domain 
may vanish.
We note that the net magnetization of the whole system
depends on termination of the system, as
net magnetizations and/or electric polarizations 
of usual antiferromagnets and/or ferroelectrics
indeed depend on the termination of the systems.
In this article, we only use the supercells tiled by tetrahedrons that are not sharing sites each other
(defined in Appendix \ref{ADWs}),
for clear-cut argument.

Here we note that our tight-binding model is introduced as the simplest
model for the holes in the $J_{\rm eff}=1/2$-manifold of $R_2$Ir$_2$O$_7$.
Therefore, we need to distinguish total angular momenta from magnetic moments of the physical spins.
In this article, we use the $J_{\rm eff}=1/2$-basis throughout and show total angular momenta
as magnetic moments. The magnetic moments of the physical spins, therefore, align
in the same direction with calculated total angular momenta and have amplitude with 1/3 of that of the total angular momenta.

The cancellation of the domain magnetizations is
similar to that of a pair of spin soliton and anti-soliton sandwiching a domain of polyacetylene\cite{SSH}.
It
becomes, however, incomplete, when external electric fields, lattice strain,
defects, charged impurities, and/or doped carriers
(as we see in Fig.~\ref{fig7}(b)) exist.
This
is essentially inverse effects of magneto-strain
and/or magneto-charge responses\cite{Arima13}.

\section{Symmetric and topological properties}
\textcolor{black}{
The low-energy effective theory of the domain-wall states and the mean-field solutions for the domain walls
in the tight-binding model have been discussed so far.
Here, we show general properties of the domain-wall states that do not depend on the
details of these theories:
Symmetric properties and topological characterizations of these domain-wall states are
given below.
First, we show the symmetric properties
both in the low-energy Dirac/Weyl equations and the mean-field
solutions of the tight-binding model, which protect the degeneracy of the domain-wall states.
Then,
\textcolor{black}{in an example of}
the (111)-domain walls, we explicitly
\textcolor{black}{demonstrate the existence of}
a weak topological number
based on the translational and three-fold-rotational symmetries of the (111)-domain walls.
}
\subsection{Degeneracy protected by symmetries}
\textcolor{black}{
Both of the low-energy effective description and the microscopic mean-field solution of the domain walls
are invariant under certain symmetry operations.
For the domain walls in the low-energy effective theories, the domain walls are trivially invariant
under the inversion around the center of the domain walls, $\hat{I}$, accompanied by the time reversal operation, $\hat{\Theta}$.
}

\textcolor{black}{
\textcolor{black}{The domain walls in the microscopic mean-field solutions
have more
symmetric properties detailed below.}
The $(01\overline{1})$-domain walls in the mean-field solution are invariant
under the inversion $\hat{I}$ around \textcolor{black}{a site} on the domain boundaries with the time reversal operation $\hat{\Theta}$.
The simultaneous two-fold rotation of the lattice and the spins around the $(01\overline{1})$-axis, $\hat{C}_2^{(01\overline{1})}$,
with $\hat{\Theta}$
also leaves the $(01\overline{1})$-domain walls invariant.  
}
\textcolor{black}{
The combination of \textcolor{black}{a} two-fold rotation around the $(001)$-axis, $\hat{C}_2^{(001)}$, and $\hat{\Theta}$
leaves the $(100)$-domain walls invariant.
The two-fold rotation around the $(100)$-axis, $\hat{C}_2^{(100)}$, also keeps the the $(100)$-domain walls unchanged.
The $(111)$-domain walls are invariant under the operations of the three-fold rotation around the $(111)$-axis, $\hat{C}_3^{(111)}$,
and the inversion with the time reversal, $\hat{I}\hat{\Theta}$.
}

\textcolor{black}{The invariance under the operation of the inversion with time reversal, $\hat{I}\hat{\Theta}$, guarantees
doubly degenerated Fermi surfaces at the $(01\overline{1})$- and $(111)$-domain walls while the invariance
under the operation of $\hat{C}_2^{(001)}$$\hat{\Theta}$ guarantees the degeneracy for the $(100)$-domain walls.
In addition, the invariance under $\hat{C}_2^{(01\overline{1})}\hat{\Theta}$ ($\hat{C}_2^{(100)}$)
guarantees the degeneracy between the two L-points projected at the same momentum in the domain-wall Brillouin zones
for the $(01\overline{1})$-domain walls (the $(100)$-domain walls).
}
\subsection{Hidden weak topological invariant}
\textcolor{black}{
In addition to the symmtric properties shown above,
the topological nature of the $(111)$-domain-wall states is
characterized by the topological invariant of the zero-dimensional class A Chern insulators\cite{RyuSchnyder,Kitaev08,Wen2012}.
In other words, as detailed below,
the domain-wall states are edge states of weak 1D Chern insulators
embedded in the bulk.}

\textcolor{black}{
First, we
\textcolor{black}{take an example of}
the (1,1,1)-domain walls.
Here we introduce the momentum frame $(k_1,k_2,k_3)$ through
\eqsa{
&&\vec{k}^{T}=\left(\frac{\pi}{4a},\frac{\pi}{4a},\frac{\pi}{4a}\right)
           +k_1 \left(\frac{1}{\sqrt{6}},\frac{1}{\sqrt{6}},-\frac{2}{\sqrt{6}}\right)
           \nn
&&         +k_2 \left(-\frac{1}{\sqrt{2}},\frac{1}{\sqrt{2}},0\right)
           +k_3 \left(\frac{1}{\sqrt{3}},\frac{1}{\sqrt{3}},\frac{1}{\sqrt{3}}\right).
}
We set $(k_1,k_2)=(0,0)$ and drop the $k_1$- and $k_2$-dependence from $\hat{\mathcal{H}}_0(\vec{k})$
to concentrate on the projection of the L-point $(\pi/4a, \pi/4a, \pi/4a)$ on the $k_1 k_2$-plane.
The $k_3$-dependence is only noted as $\hat{\mathcal{H}}_0 (k_3)$ for simplicity below.
Then we define the hamiltonian that describes the sub-Hilbert space at the projection of the L-point
or along the $\Gamma$L-line of the bulk Brillouin zone as
\eqsa{
  \hat{H}_0^{(\Gamma {\rm L})}
  &=&
  \sum_{k_3} 
  \vec{\hat{c}}^{\dagger}_{k_3}
  \hat{\mathcal{H}}_0 (k_3 )
  \vec{\hat{c}}^{\ }_{k_3}
  \nn
  &=&
  \sum_{x_3,x'_3}
  \vec{\hat{c}}^{\dagger}_{x_3}
  \hat{\mathcal{H}}'_0 (x_3,x'_3 )
  \vec{\hat{c}}^{\ }_{x'_3},
}
where one-dimensional partial Fourier transformations
are employed as $\vec{\hat{c}}^{\dagger}_{x_3}=L^{-1/2}\sum_{k_3}e^{ik_3 x_3}\vec{\hat{c}}^{\dagger}_{k_3}$
\textcolor{black}{for the number of the unit cell along the (111)-direction, and the real-space
1D hamiltonian matrix $\hat{\mathcal{H}}'_0 (x_3,x'_3 )$ is introduced}.
The 1D hamiltonian $\hat{H}_0^{(\Gamma {\rm L})}$ embedded in the bulk hamiltonian $\hat{H}_0$ describes
a hidden 1D weak Chern insulator as detailed below.
}

\textcolor{black}{
We prove that the one-dimensional hamiltonian $\hat{H}_0^{(\Gamma L)}$ describes a hidden 1D weak topological insulator
characterized by a zero-dimensional class A
topological invariant.
The 1D weak topological insulator is protected by the translational symmetry along the (111)-planes,
the three-fold rotation symmetry (C$_3$-rotation) around the (111)-axis, $\hat{C}_3^{(111)}$.
}

\textcolor{black}{
The translational symmetry prohibits
scatterings among the eigenstates at the projection of the L-point $(\pi/4a, \pi/4a, \pi/4a)$
and eigenstates at other $k$-points in the $k_1 k_2$-plane.
Then, if a perturbation $\hat{\mathcal{V}}(x_3,x'_3)$
keeps the three-fold rotation symmetry around the (111)-axis and the mirror symmetry of the (111)-plane,
the deformed hamiltonian,
\eqsa{
  \hat{H}^{(\Gamma {\rm L})}=
  \sum_{x_3, x'_3}
  \vec{\hat{c}}^{\dagger}_{x_3}
  \left[
  \hat{\mathcal{H}}'_0 (x_3,x'_3 )
  +
  \hat{\mathcal{V}} (x_3,x'_3),
  \right]
  \vec{\hat{c}}^{\ }_{x'_3}
}
is characterized by a class A topological invariant at $d=0$.
Even after the introduction of the perturbation $\hat{\mathcal{V}} (x_3,x'_3)$,
the symmetric properties of the eigenstates under the C$_3$-rotation remain unchanged
from those of the unperturbed hamiltonian.
Therefore, we classify the eigenstates of the perturbed hamiltonian
$\hat{H}^{(\Gamma {\rm L})}$ by the symmetric
properties of the eigenstates of the unperturbed hamiltonian $\hat{H}^{(\Gamma {\rm L})}_0$ as follows. 
The eigenstates of $\hat{\mathcal{H}}'_0(x_3,x'_3)$
are categorized into 8 bands, which is evident in the spectrum of the Fourier-transformed
hamiltonian $\hat{\mathcal{H}}_0 (k_3)$.
By taking account the fact that the C$_3$-rotation is a discrete symmetric operation,
the symmetric property of each eigenstate is characterized by that of the eigenstates of $\hat{\mathcal{H}}_0 (k_3=0)$.
The eigenstates of $\hat{\mathcal{H}}_0 (k_3=0)$ are classified
into the four Zeemann-splitted doublets, 2E$_{\rm 1/2u}\oplus$E$_{\rm 1/2g}\oplus$E$_{\rm 3/2u}$.
}

\textcolor{black}{
There are 8$L$ eigenstates of
the perturbed hamiltonian $\hat{H}^{(\Gamma {\rm L})}$.
Out of the 8$L$ eigenstates, 2$L$ eigenstates belong to E$_{\rm 3/2u}$.
Under the presence of the all-in/all-out orders, E$_{\rm 3/2u}$ is splitted into two groups:
$L$ states belonging to E$_{\rm 3/2u}$ are located above the Fermi level and
the other $L$ states remain under the Fermi level in the bulk insulators. 
If the perturbation $\hat{\mathcal{V}} (x_3,x'_3)$ keeps the C$_3$-rotation intact,
there are no scatterings among the $2L$ states
labeled by the
E$_{\rm 3/2u}$-states
and
the other 6$L$ states,
due to differences in the \textcolor{black}{(111)-component of angular momenta,
$m_{111}$, for}
these irreducible representations.
The wave functions belonging to 2E$_{\rm 1/2u}\oplus$E$_{\rm 1/2g}$ are
transformed as,
\eqsa{
\hat{C}_3^{(111)}\ket{\Phi;m_{111}=\pm 1/2}=e^{\pm i\pi/3}\ket{\Phi;m_{111}=\pm 1/2},
}
while the wave functions belonging to E$_{\rm 3/2u}$ are transformed
under the operation of the $\hat{C}_3^{(111)}$ as
\eqsa{
\hat{C}_3^{(111)}\ket{\Phi;m_{111}=\pm 3/2}=-\ket{\Phi;m_{111}=\pm 3/2}.
}
}

\textcolor{black}{
Here we note the following fact:
If the $L$ orbitals
\textcolor{black}{with $m_{111}=+1/2$}
are occupied in the all-out phase,
the the $2L$ orbitals
\textcolor{black}{with $m_{111}=-1/2$}
are also occupied.
When the time-reversal operation is applied, the $2L$ orbitals with
\textcolor{black}{$m_{111}=+1/2$}
and the $L$ orbitals with
\textcolor{black}{$m_{111}=-1/2$}
are necessarily occupied.
In addition, the C$_3$-rotatinal symmetry prohibits the scatterings
\textcolor{black}{among orbitals with different (111)-component of the angular
momentum $m_{111}$.}}

\textcolor{black}{
To describe the structure of the spectrum at the projected L-point in detail,
matrices are defined as
\eqsa{
 \left(H_0^{(\Gamma {\rm L})}\right)_{i,j}
 =
 \bra{0}\hat{c}_{x \nu \sigma}
 \hat{H}_0^{(\Gamma {\rm L})}
 \hat{c}_{y \mu \tau}\ket{0},
}
and
\eqsa{
 \left(H^{(\Gamma {\rm L})}\right)_{i,j}
 =
 \bra{0}\hat{c}_{x \nu \sigma}
 \hat{H}^{(\Gamma {\rm L})}
 \hat{c}_{y \mu \tau}\ket{0},
}
where $i=(x,\nu,\sigma)$ and $j=(y,\mu,\tau)$.
Then the matrix representation of the unperturbed hamiltonian is diagonalized as
\eqsa{
 H_0^{(\Gamma {\rm L})}
 =
 U_{8L\times 8L}
 \left[
 \begin{array}{ccc}
 \mbox{\boldmath$D$}_{2L} & \mbox{\boldmath$0$}_{2L \times 3L}& \mbox{\boldmath$0$}_{2L \times 3L} \\
 \mbox{\boldmath$0$}_{3L \times 2L} & \mbox{\boldmath$D$}^{(+)}_{3L} & \mbox{\boldmath$0$}_{3L \times 3L} \\
 \mbox{\boldmath$0$}_{3L \times 2L} & \mbox{\boldmath$0$}_{3L \times 3L} & \mbox{\boldmath$D$}^{(-)}_{3L} \\
 \end{array}
 \right]
 U_{8L\times 8L}^{\dagger},
}
where $\mbox{\boldmath$D$}_{2L}$ and $\mbox{\boldmath$D$}^{(\pm)}_{3L}$
are $2L\times 2L$ and $3L \times 3L$ diagonal matrices, respectively.
The sub-matrices $\mbox{\boldmath$D$}^{(\pm)}_{3L}$ represent the eigenvalues of the eigenstates
that are labeled by the 2E$_{\rm 1/2u}\oplus$E$_{\rm 1/2g}$-states.
The sub-Hilbert space of the orbitals
\textcolor{black}{with $m_{111}=+1/2$ ($m_{111}=-1/2$)}
are represented by the sub-matrix $\mbox{\boldmath$D$}^{(+)}_{3L}$ ($\mbox{\boldmath$D$}^{(-)}_{3L}$).
The C$_3$-rotational symmetry that prohibits the scattering among the three sub-Hilbert spaces represented by
$\mbox{\boldmath$D$}_{2L}$ and $\mbox{\boldmath$D$}^{(\pm)}_{3L}$
leads to an important consequence: 
The unitary matrix $U_{8L\times 8L}$ transforms the perturbed hamiltonian matrix $H^{(\Gamma {\rm L})}$
into the block-diagonalized form as
\eqsa{
 H_0^{(\Gamma {\rm L})}
 =
 U_{8L\times 8L}
 \left[
 \begin{array}{ccc}
 \mbox{\boldmath$M$}_{2L} & \mbox{\boldmath$0$}_{2L \times 3L}& \mbox{\boldmath$0$}_{2L \times 3L} \\
 \mbox{\boldmath$0$}_{3L \times 2L} & \mbox{\boldmath$M$}^{(+)}_{3L} & \mbox{\boldmath$0$}_{3L \times 3L} \\
 \mbox{\boldmath$0$}_{3L \times 2L} & \mbox{\boldmath$0$}_{3L \times 3L} & \mbox{\boldmath$M$}^{(-)}_{3L} \\
 \end{array}
 \right]
 U_{8L\times 8L}^{\dagger}.
}
Here we call the number of the occupied
orbitals that belong to $\mbox{\boldmath$M$}^{(\pm)}_{3L}$ in the all-out phase as $m_{3L}^{(\pm)}$, respectively.
Then, if the system remains gapped,
the set of the numbers of the occupied orbitals, $(m_{3L}^{(+)},m_{3L}^{(-)})$, is invariant under any perturbation
that keeps the C$_3$-rotational symmetry around the $(111)$-axis.
}

\textcolor{black}{
Therefore, the magnetic domain walls that keep the C$_3$-rotational symmetry around the (111)-axis intact,
the numbers of the occupied orbitals
$(m_{3L}^{(+)},m_{3L}^{(-)})$ give a zero-dimensional topological invariant.
If the filling of the system is kept at halffilling, the conservation of the electrons leads to $m_{3L}^{(+)}+m_{3L}^{(-)}=3L$.
Therefore, one of them gives us a $\mbox{\boldmath$Z$}$ topological invariant classified in the zero-dimensional class A\cite{Kitaev08,Wen2012}.
Here, we note that the all-in/all-out ordered phases of the pyrochlore iridium oxides do not possess
the chiral, particle-hole, and time reversal symmetries by themselves.  
}

\textcolor{black}{The robust domain-wall states must exist
when
the weak topological invariant changes
at the domain walls. The changes in the invariant indeed occur.
If we remind that the eigenstates classified by E$_{1/2g}$ are splitted into two groups under the presence of the all-in/all-out
orders, the zero-dimensional topological invariant $m_{3L}^{(+)}$ necessarily changes from $2L$ to $L$, or $L$ to $2L$
across the domain walls.}

\textcolor{black}{The interchanges in the occupation of $m_{111}=+1/2$-states and $m_{111}=-1/2$-states
correspond to the switching of the location of the 0-th Landau levels from the all-out domains to
the all-in domains.
Furthermore, as detailed in Appendix.E, the eigenstates
belonging to E$_{\rm 1/2 g}$ with $m_{111}=-1/2$ indeed
participate in the wave function of the 0-th Landau level.}

\begin{figure}[htb]
\centering
\includegraphics[width=9.0cm]{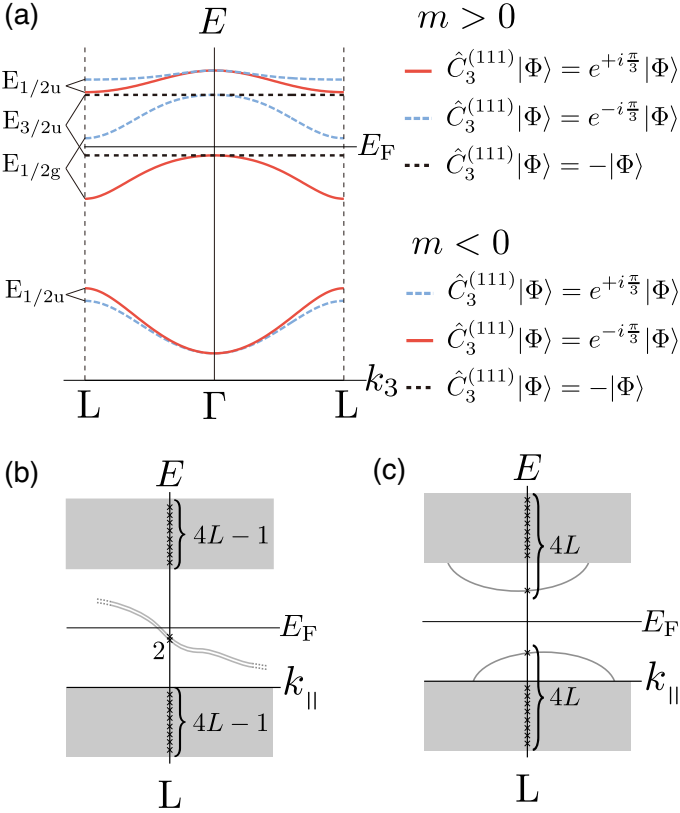}
\caption{
(a) Classification of the eigenstates at the projected L-point.
For illustrative purpose, the eigenstates are shown under the translational invariance along the (111)-axis
or the $\Gamma$L-line.
These eigenstates are invariant under the three-fold rotation $\hat{C}_3^{(111)}$
and, thus, labeled by the eigenvalues of the operator $\hat{C}_3^{(111)}$.
Here the eigenstates are $e^{\pm i\frac{\pi}{3}}$ and $-1$.
For the all-out domain ($m>0$), the red solid (blue dotted) curves
represent the eigenstates with the eigenvalue $e^{+i\frac{\pi}{3}}$ ($e^{-i\frac{\pi}{3}}$).
After applying the time reversal operation, namely in the all-in domain ($m<0$),
these eigenstates are switched as
the eigenvalue $e^{-i\frac{\pi}{3}}$ ($e^{+i\frac{\pi}{3}}$)
corresponds to 
the red solid (blue dashed) curves.
(b) Schematic energy spectrum along
momentum $k_{\parallel}$ that is perpendicular to $k_3$, 
with the protected two-fold degeneracy of the domain-wall states.
(c) Schematic possible energy spectrum along $k_{\parallel}$
without the degeneracy of the domain-wall states.
\label{figLGL}}
\end{figure}
\subsection{Gapless excitations at domain-wall states}
\textcolor{black}{
The 1D weak Chern insulators embedded in the bulk all-in/all-out ordered phases
guarantee the existence of the ingap states.
The robust ingap states, however, do not necessarily
lead to gapless quasi-particle excitations
at the domain walls.
Below, we
show
that the metallic domain-wall states
are guaranteed by
the degeneracy of the domain-wall states
protected by the symmetries of the domain walls introduced in Sec.VIIIA.}

\textcolor{black}{The appearance of the protected ingap states
due to interchange of the eigenstates above and below the Fermi level
imposes constraint on the number of eigenstates belonging to
the conduction, valence, and ingap states: At the projected L-point,
there are
$4L-\ell$ conduction and valence states while there are
$2\ell$ ingap states, where $\ell$ is some integer.}
\textcolor{black}{The low-energy effective 1D Dirac and 3D Weyl equations give
us the precise value of $\ell$, which is invariant due to the topological protection
under any perturbations that keep the symmetry of the domain walls.
For the (111)-domain walls, the number of the ingap states is given as $\ell=2$.}

\textcolor{black}{The additional symmetry of the domain walls, namely, the invariance
under $\hat{I}\hat{\Theta}$ for the (111)-domain walls,
protects the two-fold degeneracy of the ingap states.
Even away from the projected L-point, the ingap-state degeneracy is kept
as illustrated in Fig.8(b).
In addition, the ingap states never disappear unless they are merged into the
bulk  Bloch states.}

\textcolor{black}{Then, if we assume that there are no gapless excitations,
the conduction and valence bands contain $4L^3 + L^2$ and $4L^3 - L^2$ eigenstates, respectively,
or $4L^3 - L^2$ and $4L^3 + L^2$ eigenstates, respectively, for the system with the $L^3$ unit cells.
Thus, if there are no gapless excitations at the domain walls, the domain walls
require $L^2$ electron/hole dopings, which inevitably induce macroscopic electric polarizations.
It also prohibits us to keep the system at half filling.
Therefore, the domain-wall states inevitably offer gapless quasi-particle excitations
at half filling and/or without macroscopic electric polarizations.}

\textcolor{black}{In contrast, if the symmetry that protects the degeneracy of the
domain-wall states is broken, the fully gapped states at half filling is realized
by opening gaps at the domain-wall states as schematically illustrated in Fig.8(c).
The degeneracy lifting and gap opening may indeed occur at the surface between the bulk
and vacuum, because at the surface the required symmetry ($\hat{I}\hat{\Theta}$
for the (111)-domain wall) is broken.}
\subsection{24 Weyl nodes}
\textcolor{black}{
Here we
show
that, even if the 24 Weyl nodes appear as shown in
LSDA+SO+$U$\cite{Wan11}, the weak topological invariant
is unchanged.
As already mentioned in Sec.III,
the 24 Weyl nodes can be induced by certain perturbations that keep
the bulk lattice symmetry and, at least,
the invariance of the system under $\hat{C}_3^{(111)}$.
For example, the level scheme at the L-point is
controlled by introducing third-neighbor hoppings
between two sites connected
by the real space vectors with amplitudes $2\sqrt{2}a$, such as
$(2a,0,2a)$ and its transformations under the symmetric operations
belonging to the tetrahedron point group T$_{\rm d}$.
}

\textcolor{black}{
Such perturbations shift the relative energy of 2E$_{\rm 1/2u}$, E$_{\rm 1/2g}$,
and E$_{\rm 3/2u}$ \textcolor{black}{at the L-point
while the perturbations introduce a constant energy shift at the $\Gamma$-point,
independently
of the orbital classification.}
Indeed, E$_{\rm 1/2g}$-state with
\textcolor{black}{$m_{111}=+1/2$} below the Fermi level
and E$_{\rm 1/2u}$-state
 with
\textcolor{black}{$m_{111}=+1/2$} above the Fermi level
can be forced to touch each other at the L-point in the all-out phases ($m>0$).
As detailed in Ref.\onlinecite{Wan11},
the level cross between these two states induces the 24 Weyl nodes.
When the magnetic ordered moment grows further,
the 24 Weyl nodes become gapped as also shown in Ref.\onlinecite{Wan11}.
Here, we note that, 
although the third-neighbor hoppings induce the 24 Weyl nodes,
further-neighbor hoppings are required for realization of
Weyl semi-metals.}

\textcolor{black}{Even when
the Weyl semi-metals with the 24 Weyl nodes are realized,
an important fact is led:
After the gap opening of the 24 Weyl nodes,
the number of the unoccupied
orbitals with $m_{111}=+1/2$ is unchanged
in comparison with the original energy spectrum after the pair-annihilation
of the 8 Weyl nodes (depicted in Fig.8(a)).  
Therefore, the 24 Weyl nodes do not leave any trace after they are gapped out,
in contrast to the 8 Weyl nodes that leave the edge states of the 1D weak Chern
insulators after their pair-annihilations.
Irrespective of the existence or the absence of
the 24 Weyl points, after the magnetic ordered moment grows,
the topological invariants defined by such as the number of E$_{\rm 1/2u}$-state
with $m_{111}=+1/2$ are determined from physics of the 8 Weyl points
clarified here and are preserved, even when the 8 Weyl points do not
show up near the Fermi level.
}
\section{Discussion and comparison with experiments}
\label{Section7}
Here, we discuss implications and comparisons of the present theory and experimental results observed in $R_2$Ir$_2$O$_7$.
We find consistencies between our domain-wall theory and the experimental indications:
By cooling under
magnetic fields,
magnetic domain walls are formed and pinned at their favorable impurity/disorder sites
to optimize the net magnetization along the
external magnetic fields, 
thus generate a nonzero magnetization with the difference between zero and nonzero field coolings in the experiments\cite{Matsuhira11,Ishikawa}.
From Fig.~\ref{fig7}(b), we find that
$n_{\rm ex}$ excess carriers per unit cell
induce uniform magnetization $m_0$ created by insertion of the magnetic domain walls 
roughly up to $\sim g\mu_{\rm B}n_{\rm ex}$, 
when the domain wall concentration is $n_{\rm ex}$ (namely, the averaged domain size $\sim n_{\rm ex}^{-1}$ unit cells).
A realistic value $n_{\rm ex}\sim 10^{-3}$ explains the peculiar uniform magnetization ($\sim 10^{-3}\mu_B$/unit cell) universally observed experimentally\cite{Matsuhira11}. Self doping may also spontaneously stabilize such a stable domain structure.
The smaller magnetization for polycrystals\cite{Ishikawa} is consistent because magnetic domains are wiped out
more easily than those in single crystals.
Here we note that, in contrast to concentration of magnetic domain walls, the concentration of impurities/disorders
does not necessarily depend on whether
the sample is a polycrystal or a single crystal.
The larger hysteresis for stoichiometric samples\cite{Ishikawa} is simply ascribed to stronger all-in/all-out order. 

The conduction on the domain wall becomes dominating at low temperatures after the elimination of
the bulk Weyl electrons:
Strong sample dependence\cite{Ishikawa} and hysteresis in the magnetization sweep\cite{Matsuhira13} at lowest temperatures support this view.
Our gapless electronic states are doubly degenerate and localized at the 
opposite side of the domain wall each other, which generate
mutual scatterings and cause weak but notable Anderson localization.
We note that the double degeneracy is the consequence of the gapless solutions obtained from $\hat{h}^{(+)}$ and $\hat{h}^{(-)}$.
It is an intriguing future problem how the degeneracy is lifted.
Due to the degeneracy, the domain-wall Fermi surface seems to be a paramagnetic one.
However, once its degeneracy is lifted by external magnetic fields, two Zeeman-splitted Fermi surfaces
can be {\it chiral}, for example, for $(111)$-domain walls.
Each splitted Fermi surfaces have momentum-dependent spin polarization as expected in
non-coplanar itinerant magnets, and will show geometric anomalous Hall conductivities.
A tempting explanation of the large negative magnetoresistance for Nd/Gd compounds\cite{Matsuhira13,Disseler} is the 
fluctuating ferromagnetic moment of Nd induced by $m_0$ at zero field,
which scatters carriers at domain walls similarly to the double-exchange mechanism.
It is desired further to understand them more quantitatively for a better magnetic control of the transport. 

We here further discuss about the novelty of the present domain wall excitation.
We note that the analogy of the present domain wall with the solitons in polyacetylene is
helpful as an intuitive concept but should be understood with caution:
It is impossible to straightforwardly generalize solitons in polyacetylene to
3D systems and to create 2D metallic states with these solitons,
because the basic equations are not the same.
Furthermore, the topological classification of the polyacetylene (the class BDI) is
different from the present ones classified as 1D
\textcolor{black}{weak Chern insulator}
because of the crucial differences in the symmetry and spatial dimensionality, which generates different classes of topological phase. 
  As far as we know,
\textcolor{black}{magnetic domain walls have never been clarified
in the light of
the one-dimensional (weak) Chern insulators, possibly embedded in the seemingly trivial insulators.}
The possibility of such an insulator has never been pointed out in theoretical and experimental studies on magnetism of not only
the pyrochlore iridium oxides but also the magnetic domains in general.
From the viewpoint of physics of the magnetic domain and domain wall in the long history,
it has never been anticipated that it can have a topological character with metallic conduction distinct from the bulk.  Therefore, this is a new type of the magnetic excitations. 

In addition to the purely scientific advances, our result shows that the domain-wall states in pyrochlore
\textcolor{black}{iridium oxides} open a new way of building conducting 2D electron systems that are controllable through magnetic fields. Control of electric transports by external magnetic and/or electric fields has been a central idea of electronics and spintronics. The domain-wall Fermi surfaces predicted in our present work can be not only swept by external fields, after fabrication of samples, but also offer possible anomalous Hall metals with reduced back scatterings (in other words, high mobility), due to the background non-collinear all-in/all-out magnetic orders. Possible interplay between the domain-wall states and magnetic moments of magnetic ions may already be observed in Gd/Nd pyrochlore iridium oxides as a huge negative magneto-resistance.

\begin{acknowledgements}
We thank Taka-hisa Arima for fruitful discussions.
YY thanks Moyuru Kurita for enlightening discussion in the early stage
of the present study. YY also thanks Takahiro Misawa for his comments on the topological
classification of the 1D Dirac equations, and
Satoru Hayami for drawing his attention to Ref.\onlinecite{Yao}.
This work is financially supported by MEXT HPCI Strategic Programs 
for Innovative Research (SPIRE) (hp130007) and Computational Materials Science Initiative (CMSI).
Numerical calculation was partly carried out at the Supercomputer Center, 
Institute for Solid State Physics, Univ. of Tokyo. 
This work was also supported by Grant-in-Aid for 
Scientific Research {(No. 22104010, No. 22340090, and No. 23740261)} from MEXT, Japan.
\end{acknowledgements}

\appendix

\section{Derivation of low energy hamiltonian}
\label{SN1}
We study Weyl electrons 
by using 
\blue{the hamiltonian} (1), which describes
hole states of the $J_{\rm eff}=1/2$-manifold of
iridium atoms. 
After the mean-field decoupling
given in Appendix \ref{AUHF},
and using the unrestricted Hartree-Fock solution of Eq. (1)
given by the all-in/all-out magnetic order, 
we replace the $U$ term with the mean-field one with the order parameter
$m$.

We begin with a Fourier transformed form of the Hartree-Fock hamiltonian given by a $8\times 8$ hamiltonian. 
By extracting a low energy Hilbert space
around the Fermi level, it is reduced to 
a $6\times 6$ hamiltonian.
If we use $k$-group terminology at the $\Gamma$ point,
we extract ${\rm T}_{\rm 2g}\otimes {\rm E}_{1/2}$-manifold
of a double group $T_d$ from $({\rm T}_{\rm 2g}\oplus {\rm A}_{\rm 1g})\otimes {\rm E}_{1/2}$.

Next we further extract a $4\times 4$ low-energy part of the $6\times 6$ effective hamiltonian.
This corresponds to extraction of the ${\rm G}_{3/2}$-manifold (or $J_{3/2}$-manifold)
from ${\rm T}_{\rm 2g}\otimes{\rm E}_{1/2}={\rm E}_{5/2}\oplus{\rm G}_{3/2}$, in
the $k$-group terminology.

\blue{
\subsection{$8\times 8$ hamiltonian}
The Fourier transformed form of $8\times8$ hamiltonian (1) after the Hartree-Fock 
approximation is given by 
\begin{equation}
\hat{H}_0
         = \sum_{\vec{k}}\sum_{\nu=1,\cdots 4}
          \sum_{\alpha,\beta=\uparrow,\downarrow}
          \hatd{c}{\vec{k}\nu \alpha}
          \hat{\mathcal{H}}_0
          (\vec{k})
          \hatn{c}{\vec{k}\mu\beta},
\end{equation}
and $\hat{\mathcal{H}}_0=\hat{\mathcal{K}}_0+ \hat{\mathcal{Z}}_0+\hat{\mathcal{M}}_0$
with $\hat{\mathcal{K}}_0$ being the kinetic term proportional to $t$ as 
}
\begin{widetext}
\eqsa{
\hat{\mathcal{K}}_0 (\vec{k})=
-2t\hat{\sigma}_0 \left[
\begin{array}{cccc}
 0 & \cos (k_x -k_y)  & \cos (k_y -k_z) & \cos (k_z+k_x) \\
 \cos (k_x - k_y) & 0 & \cos (k_z-k_x) & \cos (k_y+ k_z) \\
 \cos (k_y - k_z) & \cos (k_z - k_x) & 0 & \cos (k_x + k_y) \\
 \cos (k_z + k_x) & \cos (k_y + k_z) & \cos (k_x + k_y) & 0 \\
\end{array}
\right],
}
$\hat{\mathcal{Z}}_0$ being the spin-orbit term proportional to $\zeta$ as 
\eqsa{
\hat{\mathcal{Z}}_0 (\vec{k})=
2i\zeta
\left[
\begin{array}{cccc}
 0 &\displaystyle +\frac{\hat{\sigma}_x + \hat{\sigma}_y}{\sqrt{2}} \cos (k_x -k_y) 
   &\displaystyle -\frac{\hat{\sigma}_y + \hat{\sigma}_z}{\sqrt{2}} \cos (k_y -k_z) 
   &\displaystyle +\frac{\hat{\sigma}_z - \hat{\sigma}_x}{\sqrt{2}}\cos (k_z+k_x) \\
 \displaystyle-\frac{\hat{\sigma}_x + \hat{\sigma}_y}{\sqrt{2}} \cos (k_x - k_y) 
 & 0 
 &\displaystyle+\frac{\hat{\sigma}_z + \hat{\sigma}_x}{\sqrt{2}}\cos (k_z-k_x) 
 &\displaystyle+\frac{\hat{\sigma}_y - \hat{\sigma}_z}{\sqrt{2}} \cos (k_y+ k_z) \\
 \displaystyle+\frac{\hat{\sigma}_y + \hat{\sigma}_z}{\sqrt{2}} \cos (k_y - k_z) 
 &\displaystyle -\frac{\hat{\sigma}_z + \hat{\sigma}_x}{\sqrt{2}} \cos (k_z - k_x) 
 & 0 
 &\displaystyle +\frac{\hat{\sigma}_x - \hat{\sigma}_y}{\sqrt{2}} \cos (k_x + k_y) \\
 \displaystyle-\frac{\hat{\sigma}_z - \hat{\sigma}_x}{\sqrt{2}}\cos (k_x + k_z) 
 &\displaystyle -\frac{\hat{\sigma}_y - \hat{\sigma}_z}{\sqrt{2}} \cos (k_y + k_z) 
 &\displaystyle -\frac{\hat{\sigma}_x - \hat{\sigma}_y}{\sqrt{2}} \cos (k_x + k_y) & 0 \\
\end{array}
\right],\nn
}
\blue{and $\hat{\mathcal{M}}_0$ being
the Hartree-Fock term of the all-in/all-out magnetic order proportional to $m$}
\eqsa{
\hat{\mathcal{M}}_0 =
\frac{m}{\sqrt{3}}
\left[
\begin{array}{cccc}
+\hat{\sigma}_x - \hat{\sigma}_y +\hat{\sigma}_z&0&0&0\\
0&-\hat{\sigma}_x + \hat{\sigma}_y +\hat{\sigma}_z&0&0\\
0&0&+\hat{\sigma}_x + \hat{\sigma}_y -\hat{\sigma}_z&0\\
0&0&0&-\hat{\sigma}_x - \hat{\sigma}_y -\hat{\sigma}_z\\
\end{array}
\right].
}
\end{widetext}

\subsection{Reduction from $8\times 8$ to $6\times 6$ hamiltonian}

\blue{
By assuming 
$|\zeta|/t, m/t, k^2  \ll 1$,}
the
leading order terms of 
the $6\times 6$ low energy effective hamiltonian
\blue{
$\hat{\mathcal{H}}_1=\hat{\mathcal{K}}_1 + \hat{\mathcal{Z}}_1+\hat{\mathcal{M}}_1$
is extracted} 
by using a projection,
\eqsa{
\hat{\mathcal{P}}_{4\times 3} =
\frac{1}{2}\left[
\begin{array}{ccc}
+1 & -1 &+1 \\
 -1 &+1 &+1 \\
+1 &+1 & -1 \\
 -1 & -1 & -1 \\
\end{array}
\right]
}
with
\begin{widetext}
\eqsa{
&&
\hat{\mathcal{K}}_1 = 
\hat{\mathcal{P}}_{4\times 3}^{T}\hat{\mathcal{K}}_0 \hat{\mathcal{P}}_{4\times 3}=
\nn
&&
-2t\hat{\sigma}_0
\left(
\cos k_x \cos k_y
\left[
\begin{array}{ccc}
-1 &  0 &  0 \\
 0 & -1 &  0 \\
 0 &  0 & +1 \\
\end{array}
\right]
+
\cos k_y\cos k_z
\left[
\begin{array}{ccc}
+1 &  0 &  0 \\
 0 & -1 &  0 \\
 0 &  0 & -1 \\
\end{array}
\right]
+
\cos k_z\cos k_x
\left[
\begin{array}{ccc}
-1 &  0 &  0 \\
 0 & +1 &  0 \\
 0 &  0 & -1 \\
\end{array}
\right]
\right)
\nn
&&
-2t\hat{\sigma}_0
\left[
\begin{array}{ccc}
 0
&\sin k_x \sin k_y
&\sin k_z \sin_x \\
 \sin k_x \sin k_y
& 0
&\sin k_y \sin k_z \\
 \sin k_z \sin k_x
&\sin k_y \sin k_z
& 0 \\ 
\end{array}
\right]
\nn
&=&
-2t\hat{\sigma}_0
\left[
\begin{array}{ccc}
-1+k_x^2 & k_x k_y & k_x k_z \\
k_y k_x & -1+k_y^2 & k_y k_z \\
k_z k_x & k_z k_y & -1+k_z^2 \\
\end{array}
\right]+\mathcal{O}(k^3)
=2t\mbox{\boldmath$1$}_6 -2 t (\vec{k}\hat{\sigma}_0)\otimes (\vec{k}\hat{\sigma}_0) +\mathcal{O}(k^3),
}
and
\eqsa{
\hat{\mathcal{Z}}_1 =
\hat{\mathcal{P}}_{4\times 3}^{T}\hat{\mathcal{Z}}_0 \hat{\mathcal{P}}_{4\times 3}=
2\sqrt{2}i\zeta
\left[
\begin{array}{ccc}
0&-\hat{\sigma}_z&+\hat{\sigma}_y\\
+\hat{\sigma}_z&0&-\hat{\sigma}_x\\
-\hat{\sigma}_y&+\hat{\sigma}_x&0\\
\end{array}
\right]
+\mathcal{O}(\zeta k^2).
}

The all-in/all-out mean field $\hat{\mathcal{M}}_0$ is projected
to the low-energy subspace as
\eqsa{
\hat{\mathcal{M}}_1 =
\hat{\mathcal{P}}_{4\times 3}^{T}\hat{\mathcal{M}}_0 \hat{\mathcal{P}}_{4\times 3}=
\frac{m}{\sqrt{3}}
\left[
\begin{array}{ccc}
0&\hat{\sigma}_z&\hat{\sigma}_y\\
\hat{\sigma}_z&0&\hat{\sigma}_x\\
\hat{\sigma}_y&\hat{\sigma}_x&0\\
\end{array}
\right].
}

\subsection{Reduction from $6 \times 6$ to $4 \times 4$}

Then, we extract a $4\times4$-hamiltonian
from \blue{$\hat{\mathcal{H}}_1 $} by using
a unitary transformation consisting of irreducible
representation ${\rm E}_{5/2}$ and ${\rm G}_{3/2}$ of the double group $T_d$,
\eqsa{
 \hat{\mathcal{U}}_{J}
 =
 \left[
 \begin{array}{ccc}
 -\frac{1}{\sqrt{3}}\hat{\sigma}_{x}&-\frac{1}{\sqrt{2}}\hat{\sigma}_{z}&+\frac{i}{\sqrt{6}}\hat{\sigma}_{y}\\
 -\frac{1}{\sqrt{3}}\hat{\sigma}_{y}&-\frac{i}{\sqrt{2}}\hat{\sigma}_{0}&-\frac{i}{\sqrt{6}}\hat{\sigma}_{x}\\
 -\frac{1}{\sqrt{3}}\hat{\sigma}_{z}&0&+\frac{2}{\sqrt{6}}\hat{\sigma}_{0}\\
 \end{array}
 \right].
}
The 1st and 2nd column of $\hat{\mathcal{U}}_{J}$ correspond to the E$_{5/2}$-irreducible representation,
and the other columns correspond to the G$_{3/2}$-irreducible representation.
Then, the kinetic term $\mathcal{K}_1$ is transformed as follows:
\eqsa{
 \hat{\mathcal{U}}_{J}^{\dagger}\hat{\mathcal{K}}_1 \hat{\mathcal{U}}_{J}
 =
 2t\mbox{\boldmath$1$}_6 -2t  
  \hat{\mathcal{U}}_{J}^{\dagger}
  (\vec{k}\hat{\sigma}_0)
  \otimes
  (\vec{k}\hat{\sigma}_0)
  \hat{\mathcal{U}}_{J}^{\ }
  =
 2t\mbox{\boldmath$1$}_6 -2t  
  \left[
  \begin{array}{cc}
\displaystyle
  \frac{1}{3}k^2\hat{\sigma}_0 & \vec{\hat{\nu}}^{\dagger}(\vec{k}) \\
  \vec{\hat{\nu}}^{\ }(\vec{k}) & \hat{\kappa}(\vec{k}) \\
  \end{array}
  \right],
}
where
\eqsa{
  \vec{\hat{\nu}}^{\dagger}(\vec{k})
  =
  \left(
  \frac{k_z k_x}{\sqrt{6}} \hat{\sigma}_0 +i\frac{k_y k_z}{\sqrt{6}} \hat{\sigma}_z -i \frac{k_x^2 - k_y^2}{\sqrt{6}}\hat{\sigma}_y
 +i\frac{2k_x k_y}{\sqrt{6}},\ 
  \frac{k^2 -3k_z^2}{3\sqrt{2}}\hat{\sigma}_z -\frac{k_z k_x}{\sqrt{2}}\hat{\sigma}_x -\frac{k_y k_z}{\sqrt{2}}\hat{\sigma}_y 
  \right),
}
and
\eqsa{
\hat{\kappa}(\vec{k})
=
\left[
\begin{array}{cc}
\displaystyle
\frac{k_x^2 + k_y^2}{2}\hat{\sigma}_0
& 
\displaystyle
-\frac{k_x^2 -k_y^2}{2\sqrt{3}}\hat{\sigma}_x -\frac{k_x k_y}{\sqrt{3}}\hat{\sigma}_y -\frac{k_z k_x}{\sqrt{3}}\hat{\sigma}_z
  +i\frac{k_y k_z}{\sqrt{3}}\hat{\sigma}_0 \\
\displaystyle
-\frac{k_x^2 -k_y^2}{2\sqrt{3}}\hat{\sigma}_x -\frac{k_x k_y}{\sqrt{3}}\hat{\sigma}_y -\frac{k_z k_x}{\sqrt{3}}\hat{\sigma}_z
 -i\frac{k_y k_z}{\sqrt{3}}\hat{\sigma}_0
&
\displaystyle
 \frac{k^2 +3k_z^2}{6}\hat{\sigma}_0 \\
\end{array}
\right].
}
\end{widetext}
The effective spin-orbit coupling and the all-in/all-out mean field
are transformed as
\blue{
\eqsa{
\tilde{\mathcal{Z}}_2 =
\hat{\mathcal{U}}_{J}^{\dagger}\hat{\mathcal{Z}}_1\hat{\mathcal{U}}_{J}
=2\sqrt{2}\zeta
\left[
\begin{array}{c|cc}
-2\times\mbox{\boldmath$1$}&\mbox{\boldmath$0$}&\mbox{\boldmath$0$}\\
\hline
\mbox{\boldmath$0$}&\mbox{\boldmath$1$}&\mbox{\boldmath$0$}\\
\mbox{\boldmath$0$}&\mbox{\boldmath$0$}&\mbox{\boldmath$1$}\\
\end{array}
\right],
}
and
\eqsa{
\tilde{\mathcal{M}}_2 =
\hat{\mathcal{U}}_{J}^{\dagger}\hat{\mathcal{M}}_1\hat{\mathcal{U}}_{J}
=
\left[
\begin{array}{c|cc}
\mbox{\boldmath$0$}&\mbox{\boldmath$0$}&\mbox{\boldmath$0$}\\
\hline
\mbox{\boldmath$0$}&\mbox{\boldmath$0$}&+im\hat{\sigma}_x\\
\mbox{\boldmath$0$}&-im\hat{\sigma}_x&\mbox{\boldmath$0$}\\
\end{array}
\right].
}
}
 
For $\zeta < 0$, by counting the number of states, it becomes
clear that the chemical potential \blue{is located} within the G$_{3/2}$-manifold to
keep the electron density at \blue{half filling}, in other words, one electron per site. 
Here, the reduction to $4\times 4$-hamiltonian \blue{$\hat{\mathcal{H}}_2\equiv \hat{\mathcal{K}}_2+\hat{\mathcal{Z}}_2+\hat{\mathcal{M}}_2$ } 
is achieved
by ignoring the off-diagonal term $\vec{\hat{\nu}}^{\dagger}(\vec{k})$,
which generates negligible corrections of $\mathcal{O}(tk^4/6\sqrt{2}|\zeta|)$.
Here, $\hat{\mathcal{K}}_2=\hat{\kappa}(\vec{k})$ while
$\hat{\mathcal{Z}}_2$ and
$\hat{\mathcal{M}}_2$
are the lower right $4\times 4$ components of $\tilde{\mathcal{Z}}_2$ and $\tilde{\mathcal{M}}_2$, respectively. 
We take the notation for the tight binding part as $\hat{h}_{4\times 4}\equiv \hat{\mathcal{K}}_2 +\hat{\mathcal{Z}}_2$, where $\hat{h}_{4\times 4}$ can be rewritten in a compact form as 
\begin{equation}
 \hat{h}_{4\times 4} (\vec{k})
 =
 \left[+2t\left(1-\frac{k^2}{3}\right)-2\sqrt{2}|\zeta|\right]
 \mbox{\boldmath$1$}_4
 -2t
 \vec{d}(\vec{k})\cdot
 \vec{\hat{\Gamma}},
\label{Luttinger}
\end{equation}
where
\eqsa{
 \vec{d}(\vec{k})^{T}=
 -
 \left(
 \frac{k_y k_z}{\sqrt{3}},
 \frac{k_z k_x}{\sqrt{3}},
 \frac{k_x k_y}{\sqrt{3}},
 \frac{k_x^2 -k_y^2}{2\sqrt{3}},
 \frac{
  3k_z^2 -k^2}{6}
 \right),\nn
}
and a vector of Dirac matrices
$
 \vec{\hat{\Gamma}}^{T}=
 \left(
 \hat{\Gamma}^1 ,
 \hat{\Gamma}^2 ,
 \hat{\Gamma}^3 ,
 \hat{\Gamma}^4 ,
 \hat{\Gamma}^5
 \right)
$,
which give time-reversal-symmetric terms for the Hilbert space of
the $4\times 4$ hamiltonian.
The above hamiltonian is nothing but a variation of the Luttinger hamiltonian.
Dirac matrices used here are defined as follows:
\eqsa{
  \hat{\Gamma}^1
  &=&
  \left[
  \begin{array}{cc}
  0 & -i\hat{\sigma}_0 \\
  +i\hat{\sigma}_0 & 0 \\ 
  \end{array}
  \right],\\
  \hat{\Gamma}^2
  &=&
  \left[
  \begin{array}{cc}
  0 & +\hat{\sigma}_z \\
  +\hat{\sigma}_z & 0 \\ 
  \end{array}
  \right],\\
  \hat{\Gamma}^3
  &=&
  \left[
  \begin{array}{cc}
  0 & +\hat{\sigma}_y \\
  +\hat{\sigma}_y & 0 \\ 
  \end{array}
  \right],\\
  \hat{\Gamma}^4
  &=&
  \left[
  \begin{array}{cc}
  0 & +\hat{\sigma}_x \\
  +\hat{\sigma}_x & 0 \\ 
  \end{array}
  \right],\\
  \hat{\Gamma}^5
  &=&
  \left[
  \begin{array}{cc}
  +\hat{\sigma}_0 & 0 \\
  0 &-\hat{\sigma}_0 \\ 
  \end{array}
  \right],
}
where $\hat{\sigma}_0$ is the two dimensional identity matrix and
$\hat{\sigma}_a$ ($a=x,y,z$) are the Pauli matrices.
From these 5 Dirac matrices, the other Dirac matrices representing time-reversal symmetry-breaking perturbations are
defined as,
\eqsa{
\hat{\Gamma}^{ab}=\left[\hat{\Gamma}^a , \hat{\Gamma}^b \right]/2i,
}
where, for example, the all-in/all-out magnetic order parameter is represented by $\hat{\mathcal{M}}_2=m\hat{\Gamma}^{54}$.

\section{Green's function}
\label{SN2}
By calculating Green's functions discussed below,
electronic spectra for the $4\times 4$ hamiltonian become
accessible.
The Green's function $\hat{G}_{4\times 4}$ for the Bloch hamiltonian $\hat{h}_{4\times 4}$
is defined by
\eqsa{
 \hat{G}_{4\times 4}(\vec{k},\omega)^{-1}
 &=&
 (\omega+\mu)\hat{\sigma}_0\otimes\hat{\tau}_0 - \hat{h}_{4\times 4}(\vec{k})-m\hat{\Gamma}^{54}
 \nn
 &=&
 \varpi (\vec{k},\omega)
 \mbox{\boldmath$1$}_4
 +2t\vec{d}(\vec{k})\cdot\vec{\hat{\Gamma}} -m\hat{\Gamma}^{54},\nn
\label{Ginv}
}
where
$\varpi (\vec{k},\omega)=\omega+\mu-2t+2\sqrt{2}|\zeta|+2tk^2 /3$.

Inverting the right-hand side of Eq. (\ref{Ginv}),
we obtain the Green's function as follows,
\begin{widetext}
\eqsa{
 &&\hat{G}_{4\times 4}
 =
 \left[
 \varpi\mbox{\boldmath$1$}_4 -2t\vec{d}
 \cdot\vec{\hat{\Gamma}}+m\hat{\Gamma}^{54}
 \right]
 \frac{
 \left[
 \varpi^2 -4t^2 
 |\vec{d}|^2
  -m^2
 \right]\mbox{\boldmath$1$}_4
 -4mt \left[
  d_3
  \hat{\Gamma}^{21}
  +
  d_2
  \hat{\Gamma}^{13}
  +
  d_1
  \hat{\Gamma}^{32}
      \right]
 }{\left[ \varpi^2 -4t^2 
 |\vec{d}|^2
 -m^2 \right]^2 -16m^2 t^2 
 \left[
 d_1^2
 + d_2^2 + d_3^2 \right]}
 \nn
 &&=
 \left[
 \frac{
 \displaystyle
 \frac{\mbox{\boldmath$1$}_4}{2}+\frac{-2t\vec{d}
 \cdot\vec{\hat{\Gamma}}+m\hat{\Gamma}^{54}}{2E_{+}
 }}
 {\varpi - E_{+}
 }
 +
 \frac{
 \displaystyle
 \frac{\mbox{\boldmath$1$}_4}{2}-\frac{-2t\vec{d}
 \cdot\vec{\hat{\Gamma}}+m\hat{\Gamma}^{54}}{2E_{+}
 }}
 {\varpi + E_{+}
 }
 \right]
 \left[
 \frac{\mbox{\boldmath$1$}_4}{2}
 -
 {\rm sign}(m)
 \frac{
 d_3
 \hat{\Gamma}^{21}
  +
  d_2
  \hat{\Gamma}^{13}
  +
  d_1
  \hat{\Gamma}^{32}
  }{2\sqrt{d_1^2 + d_2^2 + d_3^2}}
 \right]
 \nn
 &&+
 \left[
 \frac{
 \displaystyle
 \frac{\mbox{\boldmath$1$}_4}{2}+\frac{-2t\vec{d}
 \cdot\vec{\hat{\Gamma}}+m\hat{\Gamma}^{54}}{2E_{-}
 }}
 {\varpi - E_{-}
 }
 +
 \frac{
 \displaystyle
 \frac{\mbox{\boldmath$1$}_4}{2}-\frac{-2t\vec{d}
 \cdot\vec{\hat{\Gamma}}+m\hat{\Gamma}^{54}}{2E_{-}
 }}
 {\varpi + E_{-}
 }
 \right]
 \left[
 \frac{\mbox{\boldmath$1$}_4}{2}
 +
 {\rm sign}(m)
 \frac{
 d_3
 \hat{\Gamma}^{21}
  +
  d_2
  \hat{\Gamma}^{13}
  +
  d_1
  \hat{\Gamma}^{32}
  }{2\sqrt{d_1^2 + d_2^2 + d_3^2}}
 \right].
}
Here we omit $\vec{k}$- and $\omega$-dependences above and
define the functions $E_{\pm}$ as 
\eqsa{
 E_{\pm}(\vec{k})=\sqrt{4t^2 |\vec{d}(\vec{k})|^2 + m^2 \pm 4|m|t\sqrt{d_1(\vec{k})^2 + d_2(\vec{k})^2 + d_3(\vec{k})^2}}.
}
\end{widetext}
For the above calculation of the Green's function,
the following two identities are useful:
\eqsa{
|\vec{d}(\vec{k})| = |\vec{k}|^2 /3,
}
and
\eqsa{
\left(\vec{d}(\vec{k})\cdot\vec{\hat{\Gamma}}\right)^2=|\vec{d}(\vec{k})|^2.
}

\section{Domain walls}
\label{ADWs}
For each of three domain walls, namely, $(01\overline{1})$, $(111)$, and $(100)$-domain walls,
we introduce coordinate transformations given as follows.

For the $(01\overline{1})$-domain wall,
we introduce a new oblique coordinate $(X,Y,Z)$ with $(Y,Z)$ parallel to the domain wall plane
and corresponding momentum frame $(\kappa_X,\kappa_Y,\kappa_Z)$,
\eqsa{
 \vec{r}=X
 \left[
 \begin{array}{c}
  0\\
  +2a\\
  -2a\\
  \end{array}
  \right]
  + Y
 \left[
 \begin{array}{c}
  0\\
  -2a\\
  -2a\\
  \end{array}
  \right]
  +Z
 \left[
 \begin{array}{c}
  -4a\\
  +2a\\
  +2a\\
  \end{array}
  \right],
}
and
\eqsa{
 \vec{k}=\kappa_X
 \left[
 \begin{array}{c}
   0\\
   +1/4a\\
   -1/4a\\
  \end{array}
  \right]
  + \kappa_Y
 \left[
 \begin{array}{c}
  -1/4a\\
  -1/4a\\
  -1/4a\\
  \end{array}
  \right]
   +\kappa_Z
 \left[
 \begin{array}{c}
  -1/4a\\
   0\\
   0\\
  \end{array}
  \right].
  \nn
  \label{newframe}
}

For the $(111)$-domain wall,
\eqsa{
 \vec{r}=X
 \left[
 \begin{array}{c}
  -2a\\
  0\\
  -2a\\
  \end{array}
  \right]
  + Y
 \left[
 \begin{array}{c}
  -2a\\
  +2a\\
  0\\
  \end{array}
  \right]
  +Z
 \left[
 \begin{array}{c}
   0\\
  +2a\\
  -2a\\
  \end{array}
  \right],
}
and
\eqsa{
 \vec{k}=\kappa_X
 \left[
 \begin{array}{c}
   -1/4a\\
   -1/4a\\
   -1/4a\\
  \end{array}
  \right]
  + \kappa_Y
 \left[
 \begin{array}{c}
  -1/4a\\
  +1/4a\\
  +1/4a\\
  \end{array}
  \right]
   +\kappa_Z
 \left[
 \begin{array}{c}
  +1/4a\\
  +1/4a\\
  -1/4a\\
  \end{array}
  \right].\nn
}

For the $(100)$-domain wall,
\eqsa{
 \vec{r}=X
 \left[
 \begin{array}{c}
  +2a\\
  +2a\\
  0\\
  \end{array}
  \right]
  + Y
 \left[
 \begin{array}{c}
  0\\
  +4a\\
  +4a\\
  \end{array}
  \right]
  +Z
 \left[
 \begin{array}{c}
   0\\
   0\\
  +4a\\
  \end{array}
  \right],
}
and
\eqsa{
 \vec{k}=\kappa_X
 \left[
 \begin{array}{c}
   +1/2a\\
    0\\
    0\\
  \end{array}
  \right]
  + \kappa_Y
 \left[
 \begin{array}{c}
  -1/4a\\
  +1/4a\\
   0\\
  \end{array}
  \right]
   +\kappa_Z
 \left[
 \begin{array}{c}
  +1/4a\\
  -1/4a\\
  +1/4a\\
  \end{array}
  \right].\nn
}

\section{1D Dirac equations}
\label{SN3}
In this section, we derive 1D Dirac equations
that describe low energy single electron states
of the $4\times 4$ effective hamiltonian derived above.
By using the derived 1D Dirac equations, we obtain an analytic
description of domain-wall states.
For illustrative purpose, we focus on
the $(01\overline{1})$-domain wall and
domain-wall states traced back to the bulk Weyl electrons
around $\vec{k}_{\rm Weyl}=\pm\sqrt{|m|/2t}(1,1,1)^{T}$
and $\vec{k}_{\rm Weyl}=\pm\sqrt{|m|/2t}(-1,1,1)^{T}$.

Along  a symmetry axis parallel to $\vec{k}=(+1,+1,+1)$,
the $4\times 4$ hamiltonian is diagonalized by the following
unitary matrix,
\eqsa{
  \hat{\mathcal{U}}_{(1,1,1)}=
  \left[
  \begin{array}{llll}
  -ai               & -bi               & +bi               &+ai\\
  -b\vartheta       & +a\vartheta       & -a\vartheta       &+b\vartheta\\
  +b\vartheta^{\ast}& -a\vartheta^{\ast}& -a\vartheta^{\ast}&+b\vartheta^{\ast}\\
  +a                & +b                & +b                &+a\\
  \end{array}
  \right],
}
where
\eqsa{
\vartheta&=&\frac{1+i}{\sqrt{2}},\\
a&=&\frac{\sqrt{\sqrt{3}+1}}{2\cdot3^{1/4}},\\
b&=&\frac{\sqrt{\sqrt{3}-1}}{2\cdot3^{1/4}}.
}
\blue{It is useful to list up} the unitary transformation of
the matrices \textcolor{black}{$\hat{\Gamma}^i$ $(i=1,2,3,4,5)$} and $m\hat{\Gamma}^{54}$:
\begin{widetext}
\eqsa{
  \hat{\mathcal{U}}_{(1,1,1)}^{\dagger}
  \hat{\Gamma}^1
  \hat{\mathcal{U}}_{(1,1,1)}
  &=&
  \left[
  \begin{array}{cc|cc}
  +2ab(\vartheta +\vartheta^{\ast} )& -2a^2 \vartheta^{\ast} + 2b^2 \vartheta & 0& 0\\
  -2a^2 \vartheta + 2b^2 \vartheta^{\ast} & -2ab(\vartheta +\vartheta^{\ast} ) & 0& 0\\
   \hline
   0& 0& +2ab(\vartheta +\vartheta^{\ast} )& +2a^2 \vartheta - 2b^2 \vartheta^{\ast}\\
   0& 0& +2a^2 \vartheta^{\ast} - 2b^2 \vartheta & -2ab(\vartheta +\vartheta^{\ast} )\\
  \end{array}
  \right]\\
  &=&
  \left[
  \begin{array}{cc|cc}
  +1/\sqrt{3}&
  -1/\sqrt{6}+i/\sqrt{2}& 0& 0\\
  -1/\sqrt{6}-i/\sqrt{2}&
  -1/\sqrt{3}& 0& 0\\
   \hline
   0& 0& 
   +1/\sqrt{3}&
   +1/\sqrt{6}+i/\sqrt{2}\\
   0& 0& 
   +1/\sqrt{6}-i/\sqrt{2}&
   -1/\sqrt{3}\\
  \end{array}
  \right],
}
\eqsa{
  \hat{\mathcal{U}}_{(1,1,1)}^{\dagger}
  \hat{\Gamma}^2 
  \hat{\mathcal{U}}_{(1,1,1)}
  &=&
  \left[
  \begin{array}{cc|cc}
  +2ab(\vartheta +\vartheta^{\ast} )& -2a^2 \vartheta + 2b^2 \vartheta^{\ast} & 0& 0\\
  -2a^2 \vartheta^{\ast} + 2b^2 \vartheta & -2ab(\vartheta +\vartheta^{\ast} ) & 0& 0\\
   \hline
   0& 0& +2ab(\vartheta +\vartheta^{\ast} )& +2a^2 \vartheta^{\ast} - 2b^2 \vartheta\\
   0& 0& +2a^2 \vartheta - 2b^2 \vartheta^{\ast} & -2ab(\vartheta +\vartheta^{\ast} )\\
  \end{array}
  \right]\\
  &=&
  \left[
  \begin{array}{cc|cc}
  +1/\sqrt{3}&
  -1/\sqrt{6}-i/\sqrt{2}& 0& 0\\
  -1/\sqrt{6}+i/\sqrt{2}&
  -1/\sqrt{3}& 0& 0\\
   \hline
   0& 0& 
   +1/\sqrt{3}&
   +1/\sqrt{6}-i/\sqrt{2}\\
   0& 0& 
   +1/\sqrt{6}+i/\sqrt{2}&
   -1/\sqrt{3}\\
  \end{array}
  \right],
}
\eqsa{
  \hat{\mathcal{U}}_{(1,1,1)}^{\dagger}
  \hat{\Gamma}^3 
  \hat{\mathcal{U}}_{(1,1,1)}
  &=&
  \left[
  \begin{array}{cc|cc}
  +2(a^2 -b^2 )& +4ab & 0& 0\\
  +4ab & -2(a^2 -b^2 )& 0& 0\\
   \hline
   0& 0& +2(a^2 -b^2 )& -4ab\\
   0& 0& -4ab & -2(a^2 -b^2 )\\
  \end{array}
  \right]\\
  &=&
  \left[
  \begin{array}{cc|cc}
  +1/\sqrt{3}&
  +2/\sqrt{6}& 0& 0\\
  +2/\sqrt{6} &
  -1/\sqrt{3}& 0& 0\\
   \hline
   0& 0& 
   +1/\sqrt{3}& 
   -2/\sqrt{6}\\
   0& 0&
   -2/\sqrt{6}&
   -1/\sqrt{3}\\
  \end{array}
  \right],
}
\end{widetext}
\eqsa{
  \hat{\mathcal{U}}_{(1,1,1)}^{\dagger}
  \hat{\Gamma}^4
  \hat{\mathcal{U}}_{(1,1,1)}
  =
  \left[
  \begin{array}{c|cc|c}
   0& 0& 0&+i\\
   \hline
   0& 0&+i& 0\\
   0&-i& 0& 0\\
   \hline
  -i& 0& 0& 0\\
  \end{array}
  \right],
}
\eqsa{
  \hat{\mathcal{U}}_{(1,1,1)}^{\dagger}
  \hat{\Gamma}^5 \hat{\mathcal{U}}_{(1,1,1)}
  =
  \left[
  \begin{array}{c|cc|c}
   0& 0& 0&-1\\
   \hline
   0& 0&-1& 0\\
   0&-1& 0& 0\\
   \hline
  -1& 0& 0& 0\\
  \end{array}
  \right],
}
and
\eqsa{
&& +m
  \hat{\mathcal{U}}_{(1,1,1)}^{\dagger}
  \left[
   \begin{array}{cc}
   0 & -i\hat{\sigma}_x \\
   +i\hat{\sigma}_x & 0 \\
   \end{array}
   \right]
\hat{\mathcal{U}}_{(1,1,1)}
\nn
&& =
  \left[
  \begin{array}{cccc}
  +m & 0 & 0 & 0 \\
   0 &+m & 0 & 0 \\
   0 & 0 &-m & 0 \\
   0 & 0 & 0 &-m \\
  \end{array}
  \right].
}

\blue{First we consider the case of $\vec{k}_{\rm Weyl}=\pm\sqrt{|m|/2t}(1,1,1)^{T}$.
For the $(01\overline{1})$-domain wall,}
we introduce a new momentum frame $\vec{\kappa}=(\kappa_{X},\kappa_{Y},\kappa_{Z})^{T}$ as
\eqsa{
  \vec{k}&=&\frac{\kappa_{X}}{\pi}\vec{G}_{X}+\frac{\kappa_{Y}}{\pi}\vec{G}_{Y}+\frac{\kappa_{Z}}{\pi}\vec{G}_{Z}\nn
         &=&(-\kappa_{Y}-\kappa_{Z},+\kappa_{X}-\kappa_{Y},-\kappa_{X}-\kappa_{Y})^{T}.
}
Then the vector $\vec{d}(\vec{k})=(d_1, d_2, d_3, d_4, d_5)^{T}$ is transformed as
\eqsa{
 d_1 &=& \frac{-1}{\sqrt{3}}\left(-\kappa_{X}^{2}+\kappa_{Y}^{2}\right),\\
 d_2 &=& \frac{-1}{\sqrt{3}}\left\{+(\kappa_{Y}+\kappa_{Z})\kappa_{X}+(\kappa_{Y}+\kappa_{Z})\kappa_{Y}\right\}, \\
 d_3 &=& \frac{-1}{\sqrt{3}}\left\{-(\kappa_{Y}+\kappa_{Z})\kappa_{X}+(\kappa_{Y}+\kappa_{Z})\kappa_{Y}\right\}, \\
 d_4 &=& \frac{-1}{2\sqrt{3}}\left(-\kappa_{X}^2 + 2 \kappa_{Y}\kappa_{X} +\kappa_{Z}^2 +2\kappa_{Y}\kappa_{Z}\right), \\
 d_5 &=& \frac{-1}{6}\left(\kappa_{X}^2 + 6 \kappa_{Y}\kappa_{X} -\kappa_{Z}^2 -2\kappa_{Y}\kappa_{Z}\right).
}

\blue{Since $k$-independent and diagonal terms are absorbed into the chemical potential renormalization, $\hat{\mathcal{H}_2}=\hat{h}_{4\times 4}(\vec{k})+\hat{\mathcal{M}_2}$ 
may be rewritten after the unitary transformation $\hat{\mathcal{U}}_{(1,1,1)}$ as }
\eqsa{
&& \hat{\mathcal{U}}_{(1,1,1)}^{\dagger}
  \hat{h}_{4\times 4} \hat{\mathcal{U}}_{(1,1,1)} \nn
&=&-\frac{2}{3}tk^2\hat{\sigma}_0 \otimes \hat{\tau}_0 -2t\hat{\mathcal{U}}_{(1,1,1)}^{\dagger}\vec{d}\cdot\vec{\hat{\Gamma}}\hat{\mathcal{U}}_{(1,1,1)}
 \nn
 &=&
 -\frac{2}{3}tk^2\hat{\sigma}_0 \otimes \hat{\tau}_0
 -2t
 \left\{
 -\kappa_{Y}^2 \hat{\sigma}_{z} \otimes \hat{\tau}_{0}
 \right.
 \nn
 &&
 +\kappa_{Y}\kappa_{X}
 \left[
 \hat{\sigma}_x \otimes \hat{\tau}_x + \frac{1}{\sqrt{3}} \hat{\sigma}_x \otimes \hat{\tau}_y
 \right]
 \nn
 &&+
 \frac{\kappa_{Y}+\kappa_{Z}}{\sqrt{3}}\kappa_{X}
 \left[
 \sqrt{\frac{3}{2}}\hat{\sigma}_x \otimes \hat{\tau}_z
 -
 \frac{1}{\sqrt{2}}\hat{\sigma}_y \otimes \hat{\tau}_0
 \right]
 \nn
 &&
 +\kappa_{Z} (\kappa_{Z}+2\kappa_{Y})
 \left[
 -\frac{1}{6}\hat{\sigma}_x \otimes \hat{\tau}_{x}
 +\frac{1}{2\sqrt{3}}\hat{\sigma}_x \otimes \hat{\tau}_{y}
 \right]
 \nn
 &&
 \left.
 -\frac{1}{\sqrt{3}}\kappa_{Z}\kappa_{Y}
 \left[
 \frac{2}{\sqrt{3}}\hat{\sigma}_z \otimes \hat{\tau}_0
 +\frac{1}{\sqrt{6}}\hat{\sigma}_{x}\otimes \hat{\tau}_{z}
 +\frac{1}{\sqrt{2}}\hat{\sigma}_{y}\otimes \hat{\tau}_{0}
 \right]\right\},\nn
}
\blue{and 
\eqsa{
&& \hat{\mathcal{U}}_{(1,1,1)}^{\dagger}
  \hat{\mathcal{M}}_2 \hat{\mathcal{U}}_{(1,1,1)} 
= m\hat{\sigma}_0\otimes\hat{\tau}_z
}
}
\blue{where it is indeed diagonal at $\vec{k}_{\rm Weyl}=\pm\sqrt{|m|/2t}(1,1,1)^{T}$, which translates to $\kappa_X=\kappa_Z=0$,
and $\kappa_Y=\kappa_{0}/\sqrt{3}\equiv \pm \sqrt{|m|/2t}$.}

If $m>0$, the
2nd and 3rd components
constitute the Weyl electrons.
In other words, for $m>0$, the diagonal matrix $+2t\kappa_Y^2 \hat{\sigma}_z \otimes\hat{\tau}_0+m\hat{\sigma}_0\otimes\hat{\tau}_z$
\blue{is
\textcolor{black}{non zero}
for the 1st and  4th components and }
possibly has zero eigenvalues \blue{at $\kappa_X=\kappa_Z=0$ } only for the 2nd and 3rd components, \blue{namely at $2t\kappa_Y^2=m$.}
Then by extracting the 2nd and 3rd components, 
the $2 \times 2$ hamiltonian is obtained as
\begin{widetext}
\eqsa{
\hat{h}^{(+)}_{\Gamma \vec{k}_{\rm Weyl}}
 (\blue{\kappa_X, \delta\kappa_Y , \kappa_Z})
 &=&
-\frac{2t}{3}k^2
+
 \left[
 \begin{array}{cccc}
 0&0&0&0\\
 0&1&0&0\\
 0&0&1&0\\
 0&0&0&0\\
 \end{array}
 \right]
 \hat{\mathcal{U}}_{(1,1,1)}^{\dagger}
(-2t \vec{d}\cdot\vec{\hat{\Gamma}} + m \blue{\hat{\Gamma}^{54}})
 \hat{\mathcal{U}}_{(1,1,1)}
 \left[
 \begin{array}{cccc}
 0&0&0&0\\
 0&1&0&0\\
 0&0&1&0\\
 0&0&0&0\\
 \end{array}
 \right]
 \nn
 &=& -2t\kappa_Y^2 -\frac{4t}{3}\kappa_Z \kappa_Y -\frac{2t}{3}\kappa_Z^2
 -2t(\kappa_Y^2 + \frac{2}{3}\kappa_Z \kappa_Y )\hat{\sigma}_z
 -\frac{4t}{\sqrt{3}}\kappa_Y \kappa_X \hat{\sigma}_y
 +\frac{2t}{3}\kappa_Z (\kappa_Z + 2 \kappa_Y )\hat{\sigma}_x
 +m\hat{\sigma}_z
\nn
&\simeq &
 -2t\left(\frac{\kappa_0}{\sqrt{3}}\right)^2
 -4t\frac{\kappa_0}{\sqrt{3}}\delta \kappa_Y
 -\frac{4t}{3\sqrt{3}}\kappa_0 \kappa_Z
 -2t\left[
 \left(\frac{\kappa_0}{\sqrt{3}}\right)^2
 +2\frac{\kappa_0}{\sqrt{3}}\delta \kappa_Y
 + \frac{2}{3\sqrt{3}}\kappa_0 \kappa_Z  \right]
 \hat{\sigma}_z
 \nn
 &&
 -\frac{4t}{3}\kappa_0 \kappa_X \hat{\sigma}_y
 +\frac{4t}{3\sqrt{3}}\kappa_0 \kappa_Z \hat{\sigma}_x
 +m\hat{\sigma}_z
\label{m>0_1DDirac}
}

If $m<0$, zero eigenvalues may appear only for the 1st and 4th components. By extracting the 1st and 4th components,
the $2\times 2$ hamiltonian has the form similar to Eq. (\ref{m>0_1DDirac}) as,
\eqsa{
 &&\hat{h}^{(-)}_{\Gamma \vec{k}_{\rm Weyl}}
 (\blue{\kappa_X, \delta\kappa_Y , \kappa_Z})
 \nn
 &&=
 4t\frac{\kappa_0}{\sqrt{3}}
 \left\{
 -
 \left(\delta \kappa_Y
 +\frac{1}{3}
 \kappa_Z
 \right)\hat{\sigma}_0
 +
 \left(
 \delta \kappa_Y
 +\frac{1}{3}
  \kappa_Z
 \right)
 \hat{\sigma}_z
 -\frac{1}{\sqrt{3}}
 \hat{\sigma}_y
 \kappa_X 
 +\frac{1}{3}
 \kappa_Z \hat{\sigma}_x
 +
 \frac{\blue{m(X) +
 |m|}
 }{4t\kappa_0/\sqrt{3}}
 \hat{\sigma}_z
 \right\}.
 \label{m<0_1DDirac}
}
\end{widetext}

Then,
surface and domain wall states are obtained by solving 
Dirac hamiltonian (\ref{m>0_1DDirac}) and (\ref{m<0_1DDirac}).
For simplicity, we concentrate on a pair of the Weyl points,
$\vec{k}_{\rm Weyl}=\pm (\sqrt{|m|/2t},\sqrt{|m|/2t},\sqrt{|m|/2t})^{T}$,
and on a surface or domain perpendicular to $(0,+1,-1)$,
namely, $(01\overline{1})$-surface or domain.
In the following discussion, we take the coordination axis along $(0,+1,-1)$
as $X$-axis. 
Around these two Weyl points,
low-energy quasi-particle excitations are
described by the lowest order $\vec{k}\cdot\vec{p}$-type
hamiltonian $\hat{h}^{(+)}_{\Gamma\vec{k}_{\rm Weyl}}$ up to the linear order in $-i\partial_X$, $\delta\kappa_Y$, and $\kappa_Z$,
\eqsa{
 &&\hat{h}^{(+)}_{\Gamma \vec{k}_{\rm Weyl}}(-i\partial_X, \delta\kappa_Y, \kappa_Z)
 \nn
 &&=
 4t\frac{\kappa_0}{\sqrt{3}}
 \left\{
 -\left(\delta\kappa_Y + \frac{\kappa_Z}{3}\right)\hat{\sigma}_0
 +\frac{\kappa_Z}{3}\hat{\sigma}_x
 +\frac{i}{\sqrt{3}}\hat{\sigma}_y \partial_X
 \right.\nn
 &&
 \left.
 +
 \left[
 -\left(\delta\kappa_Y +\frac{\kappa_Z}{3}+\frac{|m|}{4t\kappa_0/\sqrt{3}}\right)
 +\frac{m(X)}{4t\kappa_0/\sqrt{3}}
 \right]\hat{\sigma}_z
 \right\},\nn
 \label{DkWeyl}
}
for $m(X) = +|m|$,
where we introduce a new momentum frame
and replace $\kappa_X$ with $-i\partial_X$.
Here the pair of Weyl points is given as $\vec{k}_{\rm Weyl}=- \kappa_0 (1/\sqrt{3},1/\sqrt{3},1/\sqrt{3})^{T}$
with $\kappa_0=\pm \sqrt{3|m|/2t}$.
%

Then the two component one-dimensional
Dirac equation,
\eqsa{
 \hat{h}^{(+)}_{\Gamma\vec{k}_{\rm Weyl}}(-i\partial_X, \delta\kappa_Y, \kappa_Z)
 \left[
 \begin{array}{c}
 \psi_1 (X)\\
 \psi_2 (X)\\
 \end{array}
 \right]
 =E
 \left[
 \begin{array}{c}
 \psi_1 (X)\\
 \psi_2 (X)\\
 \end{array}
 \right].\nn
 \label{S52}
}
gives description of
bound states on the surface or domain walls
by carefully choosing the $X$-dependent ``mass" term $m(X)$ as follows.
Here we note that the all-out (all-in) domain is described by $m(X)=+|m|$ ($m(X)=-|m|$).
We also remind the readers that, for a large enough order parameter $|m|$,
the Weyl points are annihilated in pair and the bulk system becomes a trivial
magnetic insulator with a charge excitation gap.
Note that
the mass term
\eqsa{
  m(X)=
  \left\{
  \begin{array}{ll}
  +|m| & ( X < 0) \\
  -|m| & ( 0 < X)\\ 
  \end{array}
  \right. ,
}
gives a magnetic domain wall at $X=0$, while
the mass term
\eqsa{
  m(X)=
  \left\{
  \begin{array}{ll}
  +|M| & ( X < 0) \\
  +|m| & ( 0 < X)\\ 
  \end{array}
  \right. ,
}
with $|M|\gg |m|$
mimics
a surface between a vacuum ($X<0$) and the bulk ($X>0$) at $X=0$.

Indeed, for the Weyl point with $\kappa_0 < 0$, we obtain the zero modes localized around the surface and the domain wall as
\begin{widetext}
\eqsa{
  \left[
 \begin{array}{c}
 \psi_1 (X)\\
 \psi_2 (X)\\
 \end{array}
 \right]
 =
 \left\{
 \begin{array}{lll}
%
 \frac{1}{\sqrt{\lambda_d + \Lambda_d}}
 \left[
 \begin{array}{cc}
 \frac{+1}{\sqrt{2}}\\
 \frac{+1}{\sqrt{2}}\\
 \end{array}
 \right]
 \left\{
 \begin{array}{cc}
 e^{+X/\lambda_d} & ( X < 0 ) \\
 e^{-X/\Lambda_d} & ( 0 < X ) \\
 \end{array}
 \right.&(\kappa_Z > 0, \delta \kappa_Y =0)&{\rm (domain\ wall)}\\
%
 \frac{1}{\sqrt{\lambda_s + \Lambda_s}}
 \left[
 \begin{array}{cc}
 \frac{+1}{\sqrt{2}}\\
 \frac{-1}{\sqrt{2}}\\
 \end{array}
 \right]
 \left\{
 \begin{array}{cc}
 e^{+X/\lambda_s} & ( X < 0 ) \\
 e^{-X/\Lambda_s} & ( 0 < X ) \\
 \end{array}
 \right.&(\kappa_Z < 0, \delta \kappa_Y =0 )&{\rm (surface)}\\
 \end{array}
 \right. ,
}
where
inverse penetration lengths are
$\lambda_d^{-1} = \kappa_Z / \sqrt{3} > 0$,
$\Lambda_d^{-1} = -3|m|/2\kappa_0-\kappa_Z/\sqrt{3} > 0$,
$\lambda_s^{-1} = -3(|M|-|m|)/2\kappa_0 +\kappa_Z / \sqrt{3} > 0$,
and
$\Lambda_s^{-1} = -\kappa_Z / \sqrt{3} > 0$.
\end{widetext}

Around the Weyl points $\vec{k}_{\rm Weyl}'=\pm (-\sqrt{|m|/2t},\sqrt{|m|/2t},\sqrt{|m|/2t})$,
by following similar procedure used for $\hat{h}^{(\pm)}_{\Gamma\vec{k}_{\rm Weyl}}$,
a Dirac hamiltonian describing low-energy quasiparticle excitations is obtained as 
\eqsa{
 &&\hat{h}^{(+)}_{\Gamma \vec{k}_{\rm Weyl}'}(-i\partial_X, \delta\kappa_Y, \kappa_Z)
 \nn
 &&=
 4t\frac{\kappa_0}{\sqrt{3}}
 \left\{
 -\left(\frac{\delta\kappa_Y}{3} - \frac{\delta\kappa_Z}{3}\right)\hat{\sigma}_0
 -\left(\frac{2\delta\kappa_Y}{3}+\frac{\delta\kappa_Z}{3}\right)\hat{\sigma}_x
 \right.\nn
 &&
 \left.
 +\frac{i\hat{\sigma}_y}{\sqrt{3}} \partial_X
 +
 \left[
 -\frac{\delta\kappa_Y}{3} +\frac{\delta\kappa_Z}{3}
 +\frac{m(X)-|m|}{4t\kappa_0 /\sqrt{3}}
 \right]\hat{\sigma}_z
 \right\},\nn
 \label{S56}
}
for $m(X) = +|m|$.
The zero modes for the above 2-component Dirac hamiltonian
are given as
\begin{widetext}
\eqsa{
  \left[
 \begin{array}{c}
 \psi_1 (X)\\
 \psi_2 (X)\\
 \end{array}
 \right]
 =
 \left\{
 \begin{array}{lll}
%
 \frac{1}{\sqrt{\lambda_d + \Lambda_d}}
 \left[
 \begin{array}{cc}
 \frac{+1}{\sqrt{2}}\\
 \frac{+1}{\sqrt{2}}\\
 \end{array}
 \right]
 \left\{
 \begin{array}{cc}
 e^{+X/\lambda_d} & ( X < 0 ) \\
 e^{-X/\Lambda_d} & ( 0 < X ) \\
 \end{array}
 \right.&(\delta\kappa_Z < 0, \delta \kappa_Y =0)&{\rm (domain\ wall)}\\
%
 \frac{1}{\sqrt{\lambda_s + \Lambda_s}}
 \left[
 \begin{array}{cc}
 \frac{+1}{\sqrt{2}}\\
 \frac{+1}{\sqrt{2}}\\
 \end{array}
 \right]
 \left\{
 \begin{array}{cc}
 e^{+X/\lambda_s} & ( X < 0 ) \\
 e^{-X/\Lambda_s} & ( 0 < X ) \\
 \end{array}
 \right.&(\delta\kappa_Z > 0, \delta \kappa_Y =0 )&{\rm (surface)}\\
 \end{array}
 \right. ,
}
where
inverse penetration lengths are
$\lambda_d^{-1} = -\delta\kappa_Z / \sqrt{3} > 0$,
$\Lambda_d^{-1} = -3|m|/2\kappa_0 +\delta\kappa_Z/\sqrt{3} > 0$,
$\lambda_s^{-1} = -3(|M|-|m|)/2\kappa_0 -\delta\kappa_Z / \sqrt{3} > 0$,
and
$\Lambda_s^{-1} = \delta\kappa_Z / \sqrt{3} > 0$.
\end{widetext}

Similarly, we can obtain zero-mode solutions for the Dirac equations
$\hat{h}^{(-)}_{\Gamma \vec{k}_{\rm Weyl}}\vec{\psi}=E\vec{\psi}$ and $\hat{h}^{(-)}_{\Gamma \vec{k}_{\rm Weyl}'}\vec{\psi}=E\vec{\psi}$.
These solutions are summarized in Figs. ~\ref{fig3}(b) 
and \ref{fig6}.
\begin{figure}[htb]
\centering
\includegraphics[width=6.0cm]{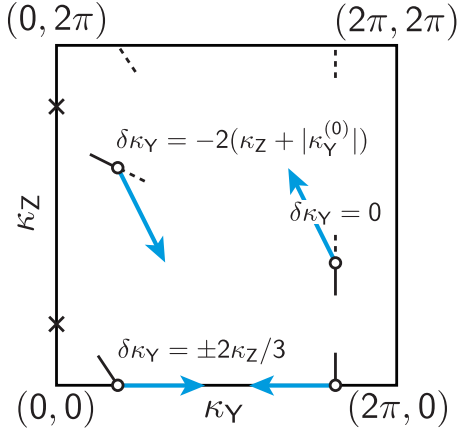}
\caption{
Summary of solutions for the Dirac equations
Loci of the domain-wall (surface) states are represented by
solid (broken) lines, which are obtained as
zero modes of the Dirac hamiltonians, $\hat{h}^{(\pm)}_{\Gamma\vec{k}_{\rm Weyl}}$,
given in Eqs. (\ref{m>0_1DDirac}), (\ref{m<0_1DDirac}), (\ref{DkWeyl}), and (\ref{S52}),
and $\hat{h}^{(\pm)}_{\Gamma\vec{k}_{\rm Weyl}'}$ (see Eq. (\ref{S56})).
Open circles and crosses illustrate the Weyl points projected to $(\kappa_Y, \kappa_Z)$-plane.
\label{figS1}}
\end{figure}

\section{Pair-annihilation of Weyl electrons at a L-point}
\label{SN4}
\begin{figure*}[htb]
\centering
\includegraphics[width=10.0cm]{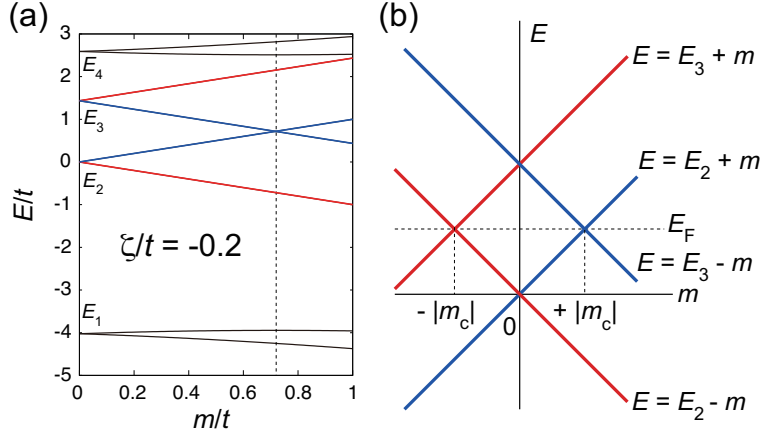}
\caption{
Level scheme of  L-point and its dependence on all-in/all-out order parameter.
(a) Level splitting at  L-point due to  finite order parameter $m$.
Here the 4 doublets are located at $E_1=-t-\sqrt{2}\zeta-\sqrt{9t^2 -6\sqrt{2}t\zeta + 6\zeta^2}$,
$E_2=0$, 
$E_3=2t+2\sqrt{2}\zeta$, 
and
$E_4=-t-\sqrt{2}\zeta+\sqrt{9t^2 -6\sqrt{2}t\zeta + 6\zeta^2}$, for $m=0$. 
The vertical broken line shows $m=|m_c|$ for which the bulk Weyl points
are annihilated in pairs.
The two blue (red) states show the states adiabatically connected to
the states that form the domain-wall state at $m=m_c$ ($m=-m_c$).
Namely, the two blue (red) lines indicate the two solutions of
the 2$\times$2 low-energy Dirac hamiltonian $\hat{h}_{\rm L}^{(+)}$ ($\hat{h}_{\rm L}^{(-)}$) (\ref{1DDirac_bulk}).
See (b) for more focused illustration.
(b) Level splitting focused on the low-energy states, around the chemical potential.
\blue{The blue states \textcolor{black}{around} the Fermi level $E_{\rm F}$ form
a domain-wall \textcolor{black}{state} confined in the side of the positive magnetization, while
the red states \textcolor{black}{around} the Fermi level are confined in the opposite $m<0$ side of the domain wall (See discussions below Eq. (\ref{solChD})).} }
\label{fig9}
\end{figure*}
To fully understand the pair-annihilation of bulk Weyl points
and formation of the closed loop of the Fermi surfaces on the domain walls,
we need to examine the structure of the original
hamiltonian, $\hat{\mathcal{K}}_0 (\vec{k})+\hat{\mathcal{Z}}_0 (\vec{k})$, around the L-point $(\pi/4a, \pi/4a, \pi/4a)$,
where the pair-annihilation occurs,
beyond the applicability of the lowest order $\vec{k}\cdot\vec{p}$-theory
developed above. 

Here we expand the $8\times 8$ Bloch hamiltonian around the L-point by
setting $\vec{k}=(\pi/4a, \pi/4a, \pi/4a)+(\delta k_x, \delta k_y, \delta k_z)$,
\begin{widetext}
\eqsa{
  \hat{\mathcal{K}}_0
  =
  -2t\hat{\sigma}_0
  \left[
  \begin{array}{cccc}
  0 & 1 & 1 & -\delta k_z-\delta k_x \\
  1 & 0 & 1 & -\delta k_y-\delta k_z \\
  1 & 1 & 0 & -\delta k_x-\delta k_y \\
  -\delta k_z-\delta k_x & -\delta k_y-\delta k_z & -\delta k_x-\delta k_y & 0\\
  \end{array}
  \right]
  +\mathcal{O}(t \delta k^2),
}
\end{widetext}
and
\eqsa{
  \hat{\mathcal{Z}}_0
  &=&
  \sqrt{2}i\zeta
  \left[
  \begin{array}{cccc}
  0 & +\hat{\sigma}_x + \hat{\sigma}_y & -\hat{\sigma}_y - \hat{\sigma}_z & 0 \\
  -\hat{\sigma}_x - \hat{\sigma}_y & 0 & \hat{\sigma}_z + \hat{\sigma}_x & 0 \\
   \hat{\sigma}_y + \hat{\sigma}_z & -\hat{\sigma}_z - \hat{\sigma}_x & 0 & 0 \\
  0 & 0 & 0 & 0 \\
  \end{array} 
  \right]
  \nn
  &+&\mathcal{O}(\zeta \delta k)
}

It is easy to determine qualitative properties of the 2-component Dirac equations
derived from the $8\times 8$ hamiltonian as follows.
First, we examine a level crossing at the L-point
under the influence of $\hat{\mathcal{M}}_0$,
which
directly corresponds to the pair-annihilation of the Weyl points.
Level scheme at the L-point for $\delta\vec{k}=\vec{0}$ and $m=0$
is given by 4 doublets,
$E=-t-\sqrt{2}\zeta\pm\sqrt{9t^2 -6\sqrt{2}t\zeta+6\zeta^2},
2t+2\sqrt{2}\zeta$, and $0$ (see Fig. ~\ref{fig9}(a)).
These doublets at the L-point are classified by the irreducible representation
of the point group D$_{\rm 3d}$, as two Kramers pairs E$_{\rm 1/2u}$
at $E=-t-\sqrt{2}\zeta\pm\sqrt{9t^2 -6\sqrt{2}t\zeta+6\zeta^2}$, one doublet E$_{\rm 3/2u}$
at $E=2t+2\sqrt{2}\zeta$, and one Kramers pair E$_{\rm 1/2g}$ at $E=0$,
if we choose the site $(0,0,0)$ in \tmag{Fig.~\ref{fig2}(b)} as the center of inversion,
which corresponds to the 7th and 8th components of the 8$\times$8 Bloch hamiltonian.
When we introduce a nonzero order parameter $m$, namely, nonzero $\hat{\mathcal{M}}_0$,
and break the time-reversal symmetry,
the
degeneracies of the four doublets are all lifted.
Then, if $2t+2\sqrt{2}\zeta > 0$ holds, the level crossing between the two states occurs \blue{at $|m|=m_c$},
one originally from the state at $E=2t+2\sqrt{2}\zeta$ and splits downward for $m\ne 0$ and
the other originally from the state at $E=0$ and splits upward for $m\ne 0$ 
\blue{(see Fig. \ref{fig9}(a))}.
The classification of the 4 doublets tells us an important fact further:
Although the all-in/all-out order parameter described by $\hat{\mathcal{M}}_0$ lifts
the degeneracy of the doublets E$_{\rm 1/2u}$ at $E=2t+2\sqrt{2}\zeta$ and the Kramers pair E$_{\rm 1/2g}$
at $E=0$, the all-in/all-out order $\hat{\mathcal{M}}_0$ does not hybridize them.
Furthermore, $\hat{\mathcal{M}}_0$ does not create matrix elements among E$_{\rm 1/2g}$ and other doublets.
Therefore, the low-energy effective theory around the pair-annihilation of 
the Weyl points only consists of these two doublets, namely, E$_{\rm 3/2u}$ and E$_{\rm 1/2g}$.

\textcolor{black}{
Here, we construct a simplified 4$\times$4 hamiltonian consisting of E$_{\rm 3/2u}$ and E$_{\rm 1/2g}$
from a full hamiltonian $\hat{\mathcal{K}}_0 + \hat{\mathcal{Z}}_0 + \hat{\mathcal{M}}_0$.
We start with four eigenfunctions
in the irreducible representations E$_{\rm 3/2u}$ and E$_{\rm 1/2g}$ at a L-point
$(\pi/4a,\pi/4a,\pi/4a)$, 
$\phi_{{\rm u \pm 3/2}}^{T}=\left[\vec{u}_{\pm}^{T},-(\hat{R}\vec{u}_{\pm})^{T},
(\hat{R}^2 \vec{u}_{\pm})^{T},0,0\right]$,
and
$\phi_{{\rm g \pm 1/2}}^{T}=\left[0,0,0,0,0,0,(1\pm\sqrt{3})(1-i)/2,1\right]/\sqrt{3\pm\sqrt{3}}$,
where we introduce a SU(2)-rotation around the (1,1,1)-axis,
$\hat{R}=\exp \left[+i(\pi/3)\cdot (1/2) \cdot (\hat{\sigma}_x + \hat{\sigma}_y
+ \hat{\sigma}_z)/\sqrt{3} \right]$,
and define
$\vec{u}_{\pm}^{T}=(\mp \sqrt{3}/2-i/2)\left[1\pm \sqrt{3},1-i\right]/\sqrt{6\pm2\sqrt{3}}$.
Then the 4$\times$4 hamiltonian up to the $\mathcal{O}(t\delta k)$ and $\mathcal{O}(t\zeta)$
is given as follows with a new momentum frame,
$k_1 = (\delta k_x + \delta k_y -2 \delta k_z)/\sqrt{6}$,
$k_2 = (-\delta k_x + \delta k_y)/\sqrt{2}$, and
$k_3= (\delta k_x + \delta k_y + \delta k_z)/\sqrt{3}$,
\eqsa{
\left[
\begin{array}{cccc}
E_{\rm g}-m &  0 &
+\frac{2}{\sqrt{6}}tk_{\rm e}^{\ast} & 
+\frac{2}{\sqrt{3}}tk_{\rm e}^{\ast} \\
0  & E_{\rm g}+m & 
-\frac{2}{\sqrt{3}}tk_{\rm e}^{\ } &
+\frac{2}{\sqrt{6}}tk_{\rm e}^{\ } \\
+\frac{2}{\sqrt{6}}tk_{\rm e}^{\ } &
-\frac{2}{\sqrt{3}}tk_{\rm e}^{\ast}&
E_{\rm u}-m & 0 \\
+\frac{2}{\sqrt{3}}tk_{\rm e}^{\ } &
+\frac{2}{\sqrt{6}}tk_{\rm e}^{\ast} &
 0 & 
E_{\rm u}+m \\
\end{array}\right],\label{H44L}
}
where
$E_{\rm g}=0$,
$E_{\rm u}=2t+2\sqrt{2}\zeta$ and $k_{\rm e}=k_1-ik_2$.}

\textcolor{black}{
By introducing external gauge fields to
the 4$\times$4 hamiltonian Eq.(\ref{H44L}),
we can easily construct Landau levels and
clarify its topological natures such as
manifestation of the chiral anomaly.
An orbital part of external magnetic fields $B$
is introduced through introducing a real space coordinate
$x_1$ corresponding to $k_1$ as 
$k_1\rightarrow -i\partial_{x_1}$,
$k_2\rightarrow k_2 -eBx_1$.
The 4$\times$4 hamiltonian Eq.(\ref{H44L}) is rewritten as
\textcolor{black}{
\eqsa{
\left[
\begin{array}{cccc}
E_{\rm g}-m &  0 &
-2it\sqrt{\frac{eB}{3}}\hat{\ell}^{-}&
-2it\sqrt{\frac{2eB}{3}}\hat{\ell}^{-}\\
0  & E_{\rm g}+m & 
-2it\sqrt{\frac{2eB}{3}}\hat{\ell}^{+}&
+2it\sqrt{\frac{eB}{3}}\hat{\ell}^{+}\\
+2it\sqrt{\frac{eB}{3}}\hat{\ell}^{+} &
+2it\sqrt{\frac{2eB}{3}}\hat{\ell}^{-}&
E_{\rm u}-m & 0 \\
+2it\sqrt{\frac{2eB}{3}}\hat{\ell}^{+}&
-2it\sqrt{\frac{eB}{3}}\hat{\ell}^{-}&
 0 & 
E_{\rm u}+m \\
\end{array}
\right],\nn\label{H44LL}
}
}
where ladder operators
\textcolor{black}{
$\hat{\ell}^{-}=+\partial_{x_1} /\sqrt{2eB} + \sqrt{eB/2} (x_1$$-$$k_2 /eB)$ and
$\hat{\ell}^{+}=-\partial_{x_1} /\sqrt{2eB} + \sqrt{eB/2} (x_1$$-$$k_2 /eB)$} are
introduced.
By using orthonormalized eigenfunctions of harmonic oscillators $\varphi_n(x)$
satisfying \textcolor{black}{$\hat{\ell}^{-}\varphi_{n}(x_1$$-$$k_2 /eB)=\sqrt{n}\varphi_{n-1}(x_1$$-$$k_2 /eB)$,
$\hat{\ell}^{+}\varphi_{n}(x_1$$-$$k_2 /eB)=\sqrt{n+1}\varphi_{n+1}(x_1$$-$$k_2 /eB)$},
and $\int dx \varphi_{n}^{\ast}(x)\varphi_{n'}(x)=\delta_{n,n'}$,
we obtain eigenvectors of Eq.(\ref{H44LL}) for Landau levels. 
}

\textcolor{black}{
Two Landau levels become important when topological properties
of the magnetic domain walls are discussed:
First one is an eigenvector
\textcolor{black}{
$\left[0,\varphi_0(x_1\right.$$-$$\left. k_2 /eB),0,0\right]^{T}$}
with an eigenvalue $E_{\rm g}+m$. 
The other is given by an eigenvector
\textcolor{black}{
$\left[0,b_1 \varphi_1(x_1\right.$$-$$k_2 /eB), b_2 \varphi_0(x_1$$-$$k_2 /eB),
c \varphi_0(x_1$$-$$\left.k_2 /eB)\right]^{T}$},
where $b_1$, $b_2$ $\rightarrow 0$ and $c \rightarrow 1$ for $|m/t|\gg 1$
with an eigenvalue approaching $E_{\rm u}+m$.
These two states are nothing but manifestation of the chiral anomaly,
in other words, 0-th Landau levels 
of Weyl nodes annihilated in pair for $E_{\rm g}+m=E_{\rm u}-m$
or $E_{\rm g}-m=E_{\rm u}+m$ at the L-points.
The asymptotic behavior of these two 0-th Landau levels
is also captured by decoupling 
$4\times 4$ hamiltonian Eq.(\ref{H44LL})
into a set of $2\times 2$ effective hamiltonians $\hat{h}_{\rm L}^{(\pm)}$.
}

\textcolor{black}{
These $2\times 2$ effective hamiltonians $\hat{h}^{(+)}_{\rm L}$
and $\hat{h}_{\rm L}^{(-)}$
consist of the \textcolor{black}{3rd and 2nd}, and, the 1st and 4th components of
the $4\times 4$ hamiltonian Eq.(\ref{H44LL}), respectively, as
\textcolor{black}{
\eqsa{
  \hat{h}^{(+)}_{\rm L}
  &=&
  \left[
  \begin{array}{cc}
  E_{\rm u}-m & -\frac{2}{\sqrt{3}}tk_{\rm e}^{\ast} \\
  -\frac{2}{\sqrt{3}}tk_{\rm e}^{ } & E_{\rm g}+m \\
  \end{array}
  \right]-\frac{E_{\rm u}+E_{\rm g}}{2}\hat{\sigma}_0,\nn
  &=&
  \left(\frac{E_{\rm u}-E_{\rm g}}{2}-m\right)\hat{\sigma}_z
  -\frac{2}{\sqrt{3}}tk_1 \hat{\sigma}_x
  +\frac{2}{\sqrt{3}}tk_2 \hat{\sigma}_y, \nn
}
and
\eqsa{
  \hat{h}^{(-)}_{\rm L}&=&
  \left[
  \begin{array}{cc}
  E_{\rm g}-m & +\frac{2}{\sqrt{3}}tk_{\rm e}^{\ast} \\
  +\frac{2}{\sqrt{3}}tk_{\rm e}^{\ } & E_{\rm u}+m \\
  \end{array}
  \right]-\frac{E_{\rm u}+E_{\rm g}}{2}\hat{\sigma}_0,\nn
  &=&
  \left(\frac{E_{\rm g}-E_{\rm u}}{2}-m\right)\hat{\sigma}_z
  +\frac{2}{\sqrt{3}}tk_1 \hat{\sigma}_x
  -\frac{2}{\sqrt{3}}tk_2 \hat{\sigma}_y,\nn 
}
after subtracting the common diagonal term
$(E_{\rm u}+E_{\rm g})\hat{\sigma}_0 /2$.}
In the following, we further derive one dimensional Dirac equations
based on $\hat{h}^{(+)}_{\rm L}$
and $\hat{h}_{\rm L}^{(-)}$.
The three-dimensional Weyl equations introduced in Sec. \ref{newV}
are also obtained as follows.
First we apply an unitary transformation,
$(\hat{\sigma}_x , \hat{\sigma}_y , \hat{\sigma}_z )
 \rightarrow
 (-\hat{\sigma}_x , +\hat{\sigma}_y , -\hat{\sigma}_z )$ for $\hat{h}^{(+)}_{\rm L}$,
and
$(\hat{\sigma}_x , \hat{\sigma}_y , \hat{\sigma}_z )
 \rightarrow
 (+\hat{\sigma}_x , -\hat{\sigma}_y , -\hat{\sigma}_z )$
for $\hat{h}^{(-)}_{\rm L}$.
Second, we rescale $(2tk_1/\sqrt{3},2tk_2/\sqrt{3},v k_3)$ to $(p_1,p_2,p_3)$ with a
constant $v$, and introduce higher order terms proportional to $p_3^2$
to reproduce pairwise annihilation of the Wey nodes.
}

Now, we focus on the $(01\overline{1})$-domain wall for illustrative purpose
and note that, up to linear orders in $\kappa_X$, $\delta \kappa_Y$, and $\kappa_Z$,
$2\times 2$-Dirac hamiltonians describing low energy physics
do not contain terms proportional to $\delta \kappa_Y$,
due to the point symmetry of the electronic band around $\vec{k}=(\pi/4a,\pi/4a,\pi/4a)$.
Here we introduce a new oblique coordinate $(\kappa_X, \delta \kappa_Y , \kappa_Z)$ through
$(\delta k_x, \delta k_y, \delta k_z)=(-\delta \kappa_Y - \kappa_Z ,+\kappa_X -\delta \kappa_Y, -\kappa_X -\delta \kappa_Y)$,
for the $(01\overline{1})$-domain wall.
It is easy to see that, along the $(111)$-direction, the band dispersion
shows a quadratic band crossing at the pair-annihilation of the Weyl points.
In other words, along the $\kappa_Y$-axis parallel to the $(111)$-direction,
the linear dispersion disappears.
Therefore, in general, the pair of the low energy $2\times 2$ Dirac hamiltonian around the L-point
is given as
\eqsa{
\hat{h}_{\rm L}^{(\pm)}&=&
  -\left[m\mp |m_{c}|\right]\hat{\sigma}_z + 
   \left( v^{(\pm)}_{Xx}\hat{\sigma}_x + v^{(\pm)}_{Xy}\hat{\sigma}_y\right)\kappa_X
  \nn
  &&+\left( v^{(\pm)}_{Zx}\hat{\sigma}_x + v^{(\pm)}_{Zy}\hat{\sigma}_y\right)\kappa_Z,
\label{1DDirac_bulk}
}
where velocities $v^{(\pm)}_{Xx}=\mp \sqrt{2}t$, $v^{(\pm)}_{Xy}=\pm2t/\sqrt{6}$,
$v^{(\pm)}_{Zx}=\pm \sqrt{2}t/3$, and $v^{(\pm)}_{Zy}=\pm2t/\sqrt{6}$
are introduced and $|m_{c}| = t +\sqrt{2}\zeta$ is the critical value
of $m$ for the pair-annihilation of the Weyl points.
Here we emphasize that the above set of the Dirac hamiltonian exploits the low energy Hilbert space
E$_{\rm 1/2u}$ $\oplus$E$_{\rm 1/2g}$.
The solution of Eq. (\ref{1DDirac_bulk}) as a function of $m$ is illustrated in Fig.~\ref{fig9}(b),
which has of course the same structure as Fig.~\ref{fig9}(a) in the low energy region.
By replacing $m$ and $\kappa$ with $m(X)$ and $-i\partial_X$ respectively,
we obtain the following 1D Dirac equation,
\begin{widetext}
\eqsa{
  \hat{h}^{(\pm)}_{\rm L}(X)=
  -\left[m(X)\mp|m_{c}|\right]\hat{\sigma}_z
  -i\widetilde{v}^{(\pm)}_X\left( \cos \varphi^{(\pm)}_X \hat{\sigma}_x + \sin \varphi^{(\pm)}_X \hat{\sigma}_y\right)\partial_X
  +\widetilde{v}^{(\pm)}_Z\left( \cos \varphi^{(\pm)}_Z\hat{\sigma}_x + \sin \varphi^{(\pm)}_Z \hat{\sigma}_y\right)\kappa_Z,
}
where velocities, $\widetilde{v}^{(\pm)}_X$ and $\widetilde{v}^{(\pm)}_Z$, and phases, $\varphi^{(\pm)}_X$ and $\varphi^{(\pm)}_Z$,
are defined through
$\widetilde{v}^{(\pm)}_X e^{i\varphi^{(\pm)}_X}=v^{(\pm)}_{Xx} + iv^{(\pm)}_{Xy}$
and 
$\widetilde{v}^{(\pm)}_Z e^{i\varphi^{(\pm)}_Z}=v^{(\pm)}_{Zx} + iv^{(\pm)}_{Zy}$.
\end{widetext}
Here the $(01\overline{1})$-domain wall with an all-out domain for $X<0$ and an all-in domain for $X>0$
is described by the following $X$-dependent mass term,
\eqsa{
  m(X)=+|m|\theta (-X) - |m|\theta (+X),
} 
which is justified from the level scheme splitting illustrated in Fig.~\ref{fig9}(b). 

Then the solution for the 1D Dirac equation $\hat{h}^{(+)}_{\rm L}(X)$,
after the pair-annihilation of the Weyl points ($|m|>|m_c|$),
\eqsa{
  \hat{h}^{(+)}_{\rm L}(X)\vec{\psi}=0\cdot\vec{\psi},
}
is given by
\eqsa{
  \vec{\psi}(X)&=&
  \frac{1}{\sqrt{\lambda_{\rm dL}+\Lambda_{\rm dL}}}
  \left[\theta (-X) e^{+X/\lambda_{\rm dL}} + \theta (+X) e^{-X/\Lambda_{\rm dL}}\right]
  \nn
  &&\times
  \left[
  \begin{array}{c}
   1/\sqrt{2}\\
   ie^{i\varphi_X}/\sqrt{2}\\
  \end{array}
  \right],
   \label{solChD}
}
where the inverse penetration lengths are $\lambda_{\rm dL}^{-1}=(|m|-|m_c|)/\widetilde{v}^{(+)}_X > 0$
and $\Lambda_{\rm dL}^{-1}=|m|/\widetilde{v}^{(+)}_X > 0$,
along the locus defined by $\kappa_Z=0$.
The above solution is mainly confined in the side of the positive magnetization $m>0$.
By solving the other 1D Dirac equation $\hat{h}^{(-)}_{\rm L}(X)\vec{\psi}=0\cdot\vec{\psi}$, 
the domain-wall state
confined in the side of the negative magnetization $m <0 $ is obtained in the same manner.

We note that, for $\kappa_Z=0$,
an operator,
\eqsa{
\hat{\Gamma}=-\sin \varphi^{(+)}_X \hat{\sigma}_x + \cos \varphi^{(+)}_X \hat{\sigma}_y ,
}
becomes
a chiral one for the above 1D Dirac equation
that satisfies the following identities,
\eqsa{
  \hat{\Gamma}^{\dagger}\hat{h}^{(+)}_{\rm L}(X)\hat{\Gamma}=-\hat{h}^{(+)}_{\rm L}(X),
}
and $\hat{\Gamma}^{\dagger}\hat{\Gamma}=1$.
Here, we also note that, in the above derivation, the chemical potential
is assumed to be pinned at the Weyl points.

Therefore, the
domain-wall zero modes
given by the solution Eq. (\ref{solChD})
is protected by the chiral operator $\hat{\Gamma}$, in other words, protected by
the chiral symmetry of $\hat{h}^{(+)}_{\rm L}(X)$ with $\kappa_Z=0$ (see Ref.2).
The classification of the topological insulators introduced in Ref.2 tells that
the 1D chiral Dirac equations derived for the low-energy physics around the L-point describe
the AIII Chern insulators.
From the above derivation, our 1D Dirac equations turn out to be chiral,
at least, around the L-point.
As long as the chemical potential is pinned at the Weyl point or the center of
the bulk gap, then the zero modes are preserved.

\section{Unrestricted Hartree-Fock treatment}
\label{AUHF}
We use the following mean-field decoupling throughout the present paper
for the unrestricted Hartree-Fock (UHF) approximation:
\eqsa{
 \hat{n}_{i\uparrow}\hat{n}_{i\downarrow}
 &\simeq&
 \left[\hatd{c}{i\uparrow},\hatd{c}{i\downarrow}\right]
 \left(
 \frac{\rho_i}{2}\hat{\sigma}_0
 -
 \frac{
 \vec{\mu}_i\cdot\vec{\hat{\sigma}}
 }{2}
 \right)
 \left[
 \begin{array}{c}
 \hatn{c}{i\uparrow}\\
 \hatn{c}{i\downarrow}\\
 \end{array}
 \right]
 \nn
 &&-
 \avrg{\hat{n}_{i\uparrow}}
 \avrg{\hat{n}_{i\downarrow}}
 +
 \avrg{\hatd{c}{i\uparrow}\hatn{c}{i\downarrow}}
 \avrg{\hatd{c}{i\downarrow}\hatn{c}{i\uparrow}},
}
where the mean fields are defined as
\eqsa{
 \rho_i &=& 
      \avrg{\hat{n}_{i\uparrow}}
      +
      \avrg{\hat{n}_{i\downarrow}},
 \\
 \mu_i^{x} &=& 
      \avrg{\hatd{c}{i\uparrow}\hatn{c}{i\downarrow}}
      +
      \avrg{\hatd{c}{i\downarrow}\hatn{c}{i\uparrow}},
 \\
 \mu_i^{y} &=& 
      -i\avrg{\hatd{c}{i\uparrow}\hatn{c}{i\downarrow}}
      +i\avrg{\hatd{c}{i\downarrow}\hatn{c}{i\uparrow}},
 \\
 \mu_i^{z} &=& 
      \avrg{\hat{n}_{i\uparrow}}
      -
      \avrg{\hat{n}_{i\downarrow}}.
}
For instance,
the all-in/all-out order
is described with the spin components of the mean-fields
$\vec{\mu}_i$ pointing in the
configuration of all-in and all-out directions\cite{WK12}.
Here a bracket $\langle \hat{O}\rangle$ means the self-consistent average of
a single particle operator $\hat{O}$.

\section{Supercells}
\label{ASCs}
For fully unrestricted Hartree-Fock calculations,
we use supercells to describe domain walls.

For the $(01\overline{1})$-domain wall calculations,
we specify the sites within the supercell as
\eqsa{
  \vec{r}_{n\ell L}=\vec{r}_n
  +\ell
 \left[
 \begin{array}{c}
  -2a\\
  +2a\\
  +0\\
  \end{array}
  \right]
 +L
 \left[
 \begin{array}{c}
   0\\
  +2a\\
  -2a\\
  \end{array}
  \right],
}
where $n$, $\ell$, and $L$ are integers, and
$\vec{r}_n$ $(n=1,2,3,4)$ is the location of the $n$-th site in the unit cell:
$\vec{r}_1=(a,0,a)^{T}$, $\vec{r}_2=(0,a,a)^{T}$, $\vec{r}_3=(a,a,0)^{T}$, and $\vec{r}_4=(0,0,0)^{T}$.

For the $(111)$-domain wall,
\eqsa{
  \vec{r}_{n 0 L}=
  \vec{r}_n
 +L
 \left[
 \begin{array}{c}
  -2a\\
   0\\
  -2a\\
  \end{array}
  \right],
}
and, for the $(100)$-domain wall,
\eqsa{
  \vec{r}_{n\ell L}=\vec{r}_n
  +\ell 
 \left[
 \begin{array}{c}
  -2a\\
  +2a\\
  +0\\
  \end{array}
  \right]
 +L
 \left[
 \begin{array}{c}
  +2a\\
  +2a\\
   0\\
  \end{array}
  \right].
}

For actual unrestricted Hartree-Fock calculations,
we chose $\ell=0,1$ and $L=0,1,\dots, 39$.
We make sharp domain walls between $L=19$ and $L=20$
as the initial conditions.
We use periodic boundary conditions parallel to the domain walls,
and open boundary conditions perpendicular to the domain walls.

\end{document}
%